\documentclass[twocolumn,3p,times]{elsarticle}
\usepackage{mathptmx, courier, pifont}
\usepackage[scaled=0.92]{helvet}
\usepackage[T1]{fontenc}
\usepackage{textcomp}
\usepackage{amsfonts}

\usepackage{epsfig}
\usepackage{amssymb}
\usepackage{graphicx}
\usepackage{amssymb}
\usepackage{graphicx}
\usepackage{rotate}
\usepackage{mathtools, cuted}
\usepackage{float}
\biboptions{numbers,sort&compress}

\begin{document}

\begin{frontmatter}

  \title{
    Systematic description of {\small COVID}-19 pandemic using exact  {\small SIR} solutions
    and Gumbel distributions 
  }

\author[a]{J.E. Amaro}
\ead{amaro@ugr.es}

\address[a]{Departamento de F\'isica At\'omica, Molecular y Nuclear
  and Instituto Carlos I de F\'isica Te\'orica y Computacional,
  Universidad de Granada, E-18071 Granada, Spain
}

\date{\today}

\begin{abstract}
  An epidemiological study of deaths is carried out in a dozen
  countries by analyzing the first wave of the {\small COVID}-19
  pandemic. These countries are among those most affected by the first
  wave, i.e. where daily-death data series may closely resemble a
  solution of the basic {\small SIR} equations. The {\small SIR}
  equations are solved parametrically using the proper time as
  parameter.  Some general properties of the {\small SIR} solutions
  are studied such as time-scaling and asymmetry. Additionally, we use
  approximations to the {\small SIR} solutions through Gumbel
  functions, which present a very similar behavior.  The parameters of
  the {\small SIR} model and the Gumbel function are extracted from
  the data and compared for the different countries. It is found that
  ten of the selected countries are very well described by the
  solutions of the {\small SIR} model, with a basic reproduction
  number between 3 and 8.
\end{abstract}

\end{frontmatter}

\section{Introduction}

Since the declaration of the {\small COVID}-19 pandemic by the World
Health Organization in March 2020, studies by the mathematical
epidemiologist community have intensified and models of various kinds
have been developed in order to provide insights and make predictions
about the spread of the disease \cite{Hui20,Tan20,Wu20,Kra20}.
Epidemiological models have been used as basic tools in epidemiology
for over a
century~\cite{Ham06,Ros08,Ros16,Ros17,Ker27,Ken57,Bar56,Bar57,Fla95}
and have been extensively used prior to the {\small COVID}-19
pandemic~\cite{Wei13,Cha14,Cha16,Rod16}.

A wide range of models have been proposed and tested in an attempt to
describe the {\small COVID}-19 data and forecast the future evolution
of the pandemic in different regions of the planet. From very simple
models \cite{Ama20,Ama21a} to numerous variants of compartmental
models based on the {\small SIR} model (susceptible, infected,
removed) have been proposed \cite{Coo20,Kud21,Pos20,Fan20}. Additional
models have been employed such as the SEIR model~\cite{Rad20}, which
adds the exposed compartment of individuals, the uncertain {\small
  SIR} model \cite{Xia21}, and others that include different
parameters that statistically describe the many factors that may
influence the pandemic dynamics. More sophisticated approaches include
deterministic or stochastic space-time models
\cite{Ama21b,nature1,linda,hakan}. Nonetheless, it has been argued
that complex models with numerous parameters may not necessarily be
advantageous without having enough data for a meaningful
validation~\cite{Rod20}.
% This
% suggests that predictions using more complex models may not necessarily  be more
% reliable compared to using a simpler model.

After going through the sixth wave in many countries, the world data
\cite{worldometer} show that a comprehensive description of the entire
time series appears to be an impossible task, since each country
presents its own characteristics (e.g. diverse lock-down and
socio-economic measures).  Therefore, in this work we proceed by
studying the first wave for those countries that present data with a
similar structure and that can unambiguously be described
mathematically using an epidemiological model of the {\small SIR}
type.

By inspection, from the recorded worldometer data \cite{worldometer}
we found that there are only nine or ten such countries (we leave
aside the case of China that has been exhaustively studied and where
the pandemic apparently died out without the need for vaccines). Among
those countries there are eight Europeans --- Spain, France, Italy,
UK, Germany, Belgium, Switzerland and Sweden --- together with Canada
and USA.  Moreover, we have also added the cases of India and Brazil.
In this work we will carry out a systematic study of the pandemic in
each of them by studying the series of cumulative deaths and daily
deaths. Our hypothesis is that deaths, $D(t)$, can be considered a
fraction of removals, $R(t)$, both cumulative and daily, and therefore
both functions follow an epidemiological curve that will differ
essentially in a normalization constant and a shift in time.

Our purpose is to investigate if data can be described with simple
epidemiological curves using the {\small SIR} model and the even
simpler Gumbel function \cite{Fur20}.  A fundamental question about
the first wave of {\small COVID}-19 is whether the lock-down
limitations had an effect in reducing the number of
deaths. Non-pharmaceutical interventions (NPI) are still under
debate. A recent meta-analysis review \cite{Her22} fails to confirm
that lock-downs have had a large, significant effect on mortality
rates.  If the daily mortality curves fit well with a basic {\small
  SIR} model, it would be interesting to conclude affirmatively or
negatively regarding the effect of NPI on them.

The structure of the paper is as follows. In Sect, II we review the
solutions of the {\small SIR} model that will be considered here.  We
describe in detail how to obtain numerical solutions as a function of
the proper time, depending on the parameters $\beta$ and the basic
reproduction number $\rho= R_0=\lambda/\beta$.  In Sect. III We
examine how well the Gumbel function fits exact {\small SIR} solutions
with only one parameter, barring normalization and a temporal shift.
In Sect IV We will discuss the time scaling of {\small SIR} solutions
and define an asymmetry parameter that depends linearly on $\rho$, and
therefore can be used to characterize the value of the basic
reproduction number from a set of data.  In Sect. V we present our
results of fits of death data with the exact {\small SIR} model and
Gumbel functions.  In Sect VI we draw our conclusions.

\section{Solutions of the {\small SIR} model and the proper time}

In this section we briefly describe the {\small SIR} model and discuss
its analytical solution in terms of, what we will call here, {\em
 proper time}, $\tau$, which is a natural variable to measure time
through the proportion of removals, where the {\small SIR} equations
have a trivial and easily interpreted solution. The real time is then
obtained by integrating the exact solution.

In the {\small SIR} model the individuals of a closed population $N$
affected by a contagious disease
are divided into three types: susceptible, $S$, infected, $I$, and
removed (recovered or dead), $R$. As functions of time, the number of
individuals in each compartment is assumed to verify the following equations
\begin{eqnarray}
\frac{dS}{dt} & = & - \frac{\lambda}{N}IS \label{ds},  \\  
\frac{dR}{dt} & = &  \beta I \label{dr}, \\
I & = & N -R -S.  \label{di}
\end{eqnarray}
In this transmission-dynamics
system 
the first equation means that the variation of susceptible individuals
decreases by infection and is proportional to the number of
susceptible and the number of infected individuals. The constant $\lambda$ 
measures the rate of infection.
The second equation describes the removal variation as
proportional to the number of infected individuals, the removal rate
being $\beta$.  By the third equation, the difference between the
total number minus the susceptible individuals minus the removed ones
must be the number of infected at each instant $ t $.

We will consider the initial values $S (0) = S_{_0} <N$ and $R (0) =
0$. 
Therefore $I (0) = N-S_{_0}> 0$. There must be a number, albeit
small, of infected in the system initially for the epidemic to
begin. For convenience below we will work with the percentages of
susceptible, infected and recovered individuals, over the total number of the
population, which are obtained by dividing by $N$:
\begin{equation}
  s(t)= \frac{S}{N}, \kern 1cm
  i(t)= \frac{I}{N}, \kern 1cm
  r(t)= \frac{R}{N},
\end{equation}
with $s(0)=s_{_0}$, $i(0)=1-s_{_0}$, and $r(0)=0$.

The proper time, $\tau$,
is defined as the temporal variable that describes naturally
the evolution of the epidemic, by counting the evolution of the
recovered individuals, $r(t)$, which is always
an increasing function with time. From
the second {\small SIR} equation in differential form, it is defined by
\begin{equation}
  dR = \beta I dt \equiv N d\tau,  \label{propertime}
\end{equation}
therefore, if we demand that $\tau = 0$ for $t = 0$, we trivially have
\begin{equation}
  \tau = r(t).
  \end{equation}

The idea of proper time is based on other equivalent approaches
described e.g. in \cite{Har14,Mil12}, where the susceptible function
$s$ is used as variable instead of $r$. The proper time is nothing
more than a change of the time variable into a more convenient one.
In our case, time is measured by counting the number of recovered (in
percent), since $r(t)$ is an increasing function, although it does not
depend linearly on time. By definition the proper time has a range
limited by
\begin{equation}
  0 \leq \tau \leq 1.
\end{equation}
From Eq. \ref{propertime} we have
\begin{equation}
  I = \frac{N}{\beta}\frac{d\tau}{dt}, \kern 1cm
  i = \frac{1}{\beta}\frac{d\tau}{dt}.
\end{equation}
Thus the change of the physical time is given by
\begin{equation} \label{time}
  dt = \frac{1}{\beta}\frac{d\tau}{i(\tau)},
\end{equation}
 where $i(\tau)$ are the infected percent expressed as a function of
 proper time.

To obtain the susceptible function note that we can write, inserting Eq. (\ref{propertime}) into Eq. (\ref{ds}) 
\begin{equation} \label{dstau}
dS = - \frac{\lambda}{N}IS dt = - \frac{\lambda}{\beta} S d\tau = -\rho S d\tau,
\end{equation}
where $\rho$ is  the so-called basic reproduction number
\begin{equation}
\rho \equiv {\cal R}_0 \equiv  \frac{\lambda}{\beta}. 
\end{equation}
The parameter $\rho$
 has here the meaning of being the decay constant of the susceptible population 
in units of proper time. 
Eq. (\ref{dstau}) is readily integrated giving
\begin{equation}
S = S_{_0}{\rm e}^{-\rho \tau},
\end{equation}
which is similar to the exponential decay law of radioactive samples as a function of proper time. Thus 
\begin{equation}
\tau_{1/2} = \frac { \ln 2}{ \rho} = \frac { \beta \ln 2 }{ \lambda} 
\end{equation}
represents the half-life in proper time units, i.e., the length of proper time
after which the susceptible population is reduced to half.

\begin{figure*}[ht]
\begin{center}
\includegraphics[width=10cm,height=10cm,angle=-0]{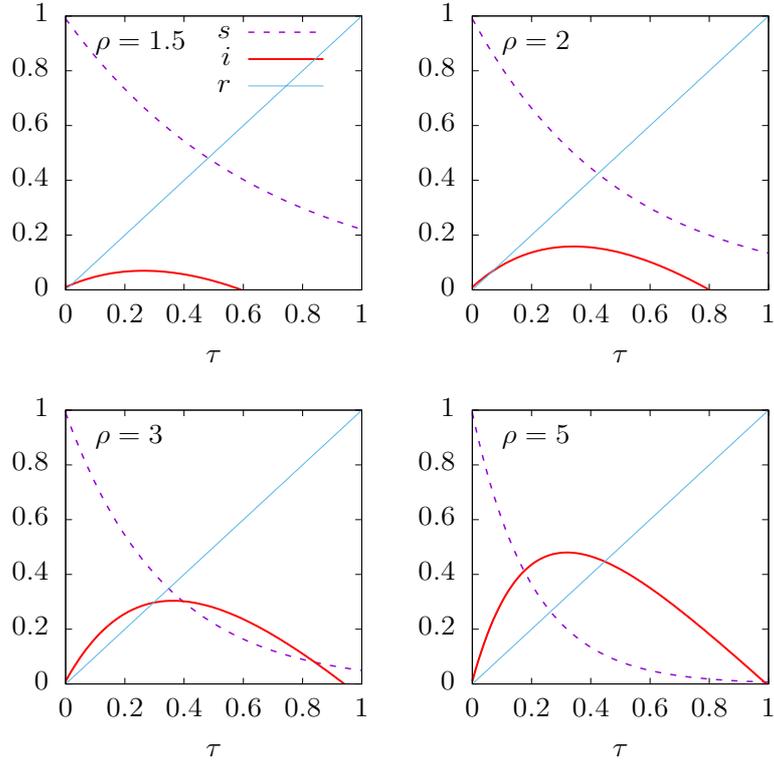}
\caption{\label{fig1}
Solution of the {\small SIR} equations as a function of proper time $\tau$ for initial susceptible $s_{_0}=0.99$, and for several values of the basic reproduction number 
$\rho={\cal R}_0$. 
 }
\end{center}
\end{figure*}

Finally, the third {\small SIR} equation (\ref{di}) gives directly the infected
population as a function of proper time
\begin{eqnarray}
I & = & N -N\tau - S_{_0}{\rm e}^{-\rho \tau}\\
i(\tau) & = & 1 -\tau - s_{_0}{\rm e}^{-\rho \tau},
\end{eqnarray}
with the condition $0 \leq i \leq 1$. For $\tau=0$ we obtain 
the initial number of infected population $i_{_0}=1-s_{_0}$. 
The end of the epidemic is reached when $i = 0$, that is, when the proper time 
verifies the transcendental equation
\begin{equation}
i(\tau_f) = 0  \Longrightarrow  \tau_f +  s_{_0}{\rm e}^{-\rho \tau_f} = 1
\end{equation} 

In Fig. \ref{fig1} we show some numerical examples. The analytical
solutions of the {\small SIR} equations are plotted as a function of proper time for
various values of the basic reproduction number $\rho=1.5,$ 2, 3 and 5. 
In all cases we
assume that $s_{_0} = 0.99$; i.e., that one percent of the population is
initially infected.

\begin{figure}
\begin{center}
\includegraphics[width=5.5cm,height=5.5cm,angle=-0]{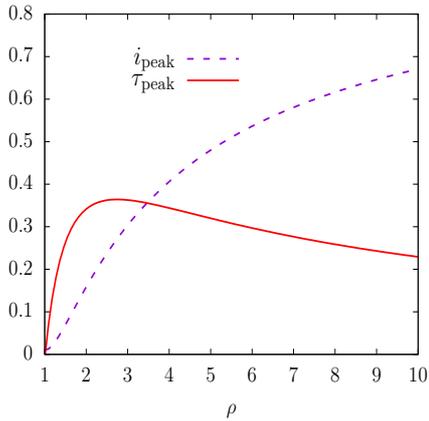}
\caption{\label{fig2}
The peak values of $i$, $\tau$ as a function of the basic reproduction number $\rho$. 
For $s_{_0}=0.99$ the peak values are almost independent of $s_{_0}$.
 }
\end{center}
\end{figure}

In Fig. \ref{fig1} we see that the number of infected individuals as a function
of $\tau$ first grows to a maximum and then decreases to zero. The
maximum of $i(\tau)$ is the peak of the epidemic.
It is reached for $di/d\tau=0$. Then
\begin{equation}
-1+\rho s_{_0}{\rm e}^{-\rho \tau} = 0,
\end{equation}
and therefore
\begin{equation}
s_{_0}{\rm e}^{-\rho \tau} = \frac{1}{\rho}.  
\end{equation}
From here the peak of the epidemic verifies
\begin{eqnarray}
  \tau_{peak} &=& \frac{\ln s_{_0}\rho}{\rho}\label{taumax} \\
s_{peak}= s(\tau_{peak}) &=& \frac{1}{\rho}\\ 
i_{peak}=i(\tau_{peak}) &=& 1- \frac{1+\ln s_{_0}\rho}{\rho}. \label{imax}
\end{eqnarray}
A condition for the existence of this maximum for $\tau_{peak}>0$, from
Eq. (\ref{taumax}), is that $s_{_0} \rho > 1$.  If we assume that, at the
very beginning of the epidemic, $s_{_0}$ is very close to 1, then it is
enough that $\rho> 1$ for the epidemic to begin to grow. In that case,
$i(\tau)$ starts growing up to a maximum reached at $\tau_{peak}$,
where it starts to decrease to zero.

In Fig. \ref{fig2} we show the peak values of the infected rate
$i_{\rm peak}(\rho)$ in the epidemic as a function of the basic reproduction
number $\rho$, for $s_{_0} = 0.99$. Since $s_{_0}$ is very close to one and
the dependence on $s_{_0}$ is logarithmic, these peak values are almost
independent on the precise value of $s_{_0} \simeq 1$.  The peak value of 
infected individuals grows with the basic reproduction number $\rho$.
When $\rho$ is very large, above 10, the logarithmic dependence on the
numerator makes the peak to grow more slowly.
For $\rho = 10$ we have $i_{\rm peak} = 0.67$, that is,
at the peak of the epidemic two thirds of the population will be
infected simultaneously.  For $\rho=50$, more than 90\% of the population will
be infected simultaneously at the peak.

In Fig. \ref{fig2} we also show the value of the proper time at the epidemic 
peak, $\tau_{\rm peak}(\rho)$.
It presents a maximum for 
\begin{eqnarray}
\frac{d\tau_{\rm peak}}{d\rho}=\frac{1-\ln s_{_0}\rho}{\rho^2}=0
\\
\ln s_{_0}\rho = 1 \Longrightarrow  \rho= \frac{e}{s_{_0}} \simeq 2.72,
\end{eqnarray}
for $s_{_0} \simeq 1$. The maximum value of $\tau_{\rm peak}$ is then
\begin{equation}
(\tau_{\rm peak})_{\rm max}= \frac{s_{_0}}{e} \simeq 0.367
\end{equation}
for $\rho = e = 2.72$.

\begin{figure*}[ht]
\begin{center}
\includegraphics[width=11cm,height=11cm,angle=-0]{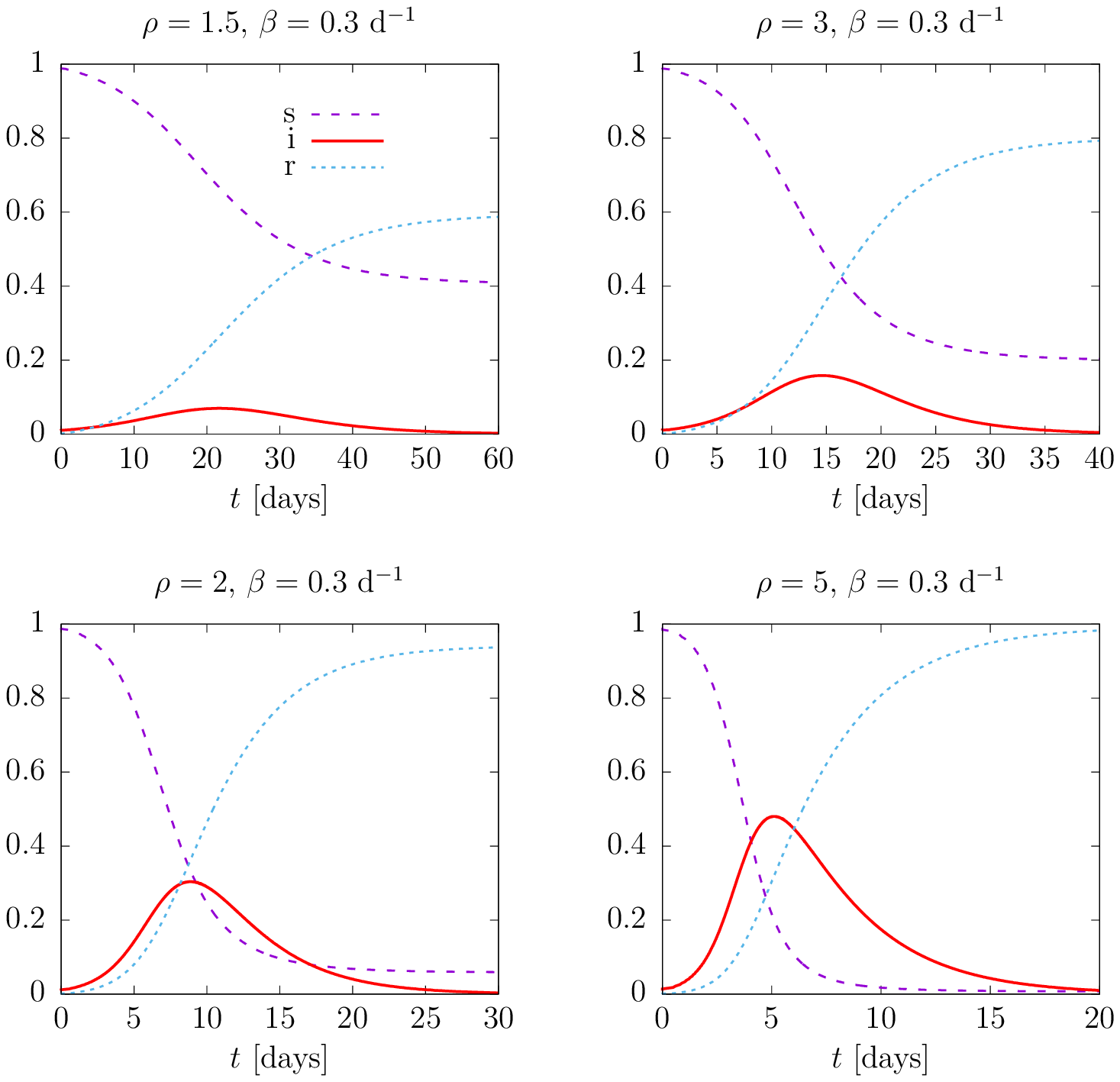}
\caption{\label{fig3}
  Solution of the {\small SIR} equations as a function of
  physical time, $t$, for initial susceptible $s_{_0}=0.99$, for
  $\beta=0.3$ d$^{-1}$, and for several values of the reproduction number
  $\rho={\cal R}_0$.  }
\end{center}
\end{figure*}

Finally, starting from the analytical solution of the {\small SIR} equations as a
function of the proper time, $ i(\tau) $, we will proceed to obtain the
solution as a function of the physical time, $t$. It is obtained from
equation (\ref{time})
by integrating between 0 and $\tau$. Assuming that
$t = 0$ for $\tau = 0$ we obtain
\begin{equation}
  t(\tau)
  = \frac{1}{\beta}\int_0^\tau \frac{d\tau}{i(\tau)}
= \frac{1}{\beta}\int_0^\tau \frac{d\tau}{1-\tau-s_{_0}{\rm e}^{-\rho\tau}}
  \end{equation}
This integral is not analytical, but it can be calculated numerically
with precision using any numerical algorithm, such as Simpson's rule
or Gaussian integration, since the integrand is quite smooth.

The final {\small SIR} solution is then obtained in parametric form by
tabulating $ t (\tau)$, $ i (\tau) $, $ \tau $ and $ s (\tau) $.
By plotting $ i (\tau) $, $ \tau $ and $ s (\tau) $
as a function of $ t (\tau) $ we obtain
the results of Fig. 3, corresponding to the exact (numerical)
solution of the {\small SIR} equations as a function of physical time, for the
same parameters of  Fig. 1 and for $\beta=0.3$ d$^{-1}$.

Note in Fig. \ref{fig3}, firstly, that the height of the maximum of
$i(t)$ coincides with that of the maximum of the analytical solution $
i(\tau) $ of Fig. 1, as it should be, since we have only made a
change of variable ---from the proper time to the physical time. Of
course, the dependence on physical time has drastically changed. For a
constant recovery rate $\beta=0.3$ d$^{-1}$,
by increasing the value of the basic reproduction
number, $\rho$, the epidemic passes more quickly and lasts less time, almost
explosively for $\rho = 5$ ---for which it lasts only twenty days---
compared to $\rho =
1.5$ ---where it lasts almost two months.

Therefore, assuming that the recovery rate, $\beta$, in an epidemic is
somewhat constant, the basic reproduction number $\rho$ largely determines
the evolution of the pandemic. The analytical solution allows
estimating the maximum number of simultaneous infections at the peak,
$i_{\rm peak}$, from this
number. This does not depend much on $s_{_0}$, as long as this number is
close to one, due to its dependence on the logarithm of $s_o\rho$.

\begin{figure*}[ht]
\begin{center}
\includegraphics[width=11cm,height=11cm,angle=-0]{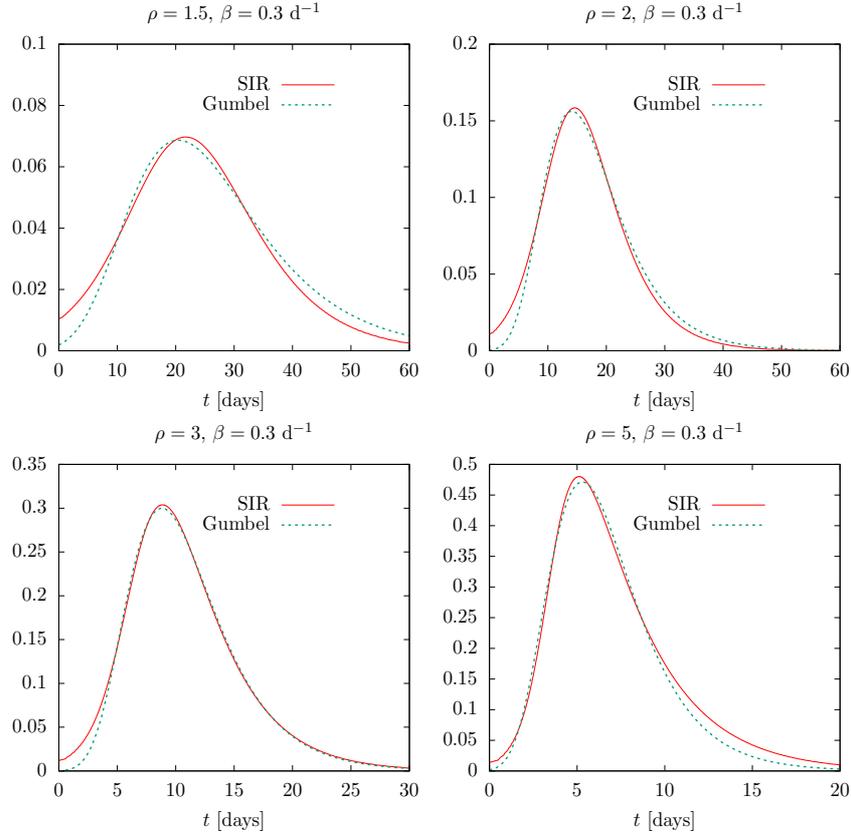}
\caption{\label{fit1}
  Gumbel distribution compared to the exact
  Solution of the {\small SIR} equations
   for $s_{_0}=0.99$,
  $\beta=0.3$ d$^{-1}$, and for several values of the reproduction number
  $\rho$.  }
\end{center}
\end{figure*}

Finally, note that the exact solution $ i (\tau)$ of the {\small SIR}
model is asymmetric. It rises very fast at first --- exponentially
actually --- and then falls more slowly.  The most explosive, severe
outbreaks occur for $\rho \geq 5$, where the end of the epidemic is
reached for $\tau_f\simeq 1$ (see Eq. (16)), thus most of the initial
susceptible individuals have been infected. For lower reproduction number the epidemic
curve becomes more symmetrical and it is less explosive.
Note that In the
first wave of the {\small COVID}-19 pandemic daily deaths rose very rapidly
and fell more slowly, indicating a high reproduction
number, so that the data could not be fitted with logistic-type
functions, which are symmetric, and a linear combination of two
logistic functions was required to fit the data
\cite{Ama20,Ama21a,Ama21b}.  In the next section we will see that some
more appropriate functions to describe this situation are the Gumbel
functions, because they have the adequate asymmetry to fit almost
correctly the solutions of {\small SIR} equations.

\section{Approximation to the {\small SIR} solutions with Gumbel functions}
%------------------------------------------------------------------

In reference \cite{Fur20}, a Gumbel distribution was used to forecast
the series of daily deaths with {\small COVID}-19 positives.  The Gumbel
distribution describes the probability of maximum (or minimum) values
from the data of many observations \cite{Gum35,Gum54}. it is not
theoretically clear why the distribution of deaths or infections
roughly recreates the distribution of maxima.
In this section we will compare the Gumbel distribution
to the exact solutions of the epidemiological {\small SIR} model. Both functions
of time presents similar asymmetry and depend on three parameters,
that can be related numerically.

\begin{table}
\begin{center}
  \begin{tabular}{cccccc}\hline\hline
   $s_{_0}$& $\rho$ & $\beta$ & $a$  & $t_0$ & $b$\\ \hline 
    0.99 & 1.5    &  0.3    & 2.04 & 20.5  & 10.9 \\
         & 2      &         & 2.67 & 14.9  & 6.29 \\
         & 3      &         & 3.06 & 8.81  & 3.75 \\
         & 5      &         & 3.12 & 5.30  & 2.44 \\\hline
  \end{tabular}
  \caption{Parameters of the Gumbel distribution fitted to the exact
    {\small SIR} model, for four different values of the basic reproduction number
    $\rho$.  }
\end{center}    
\end{table}

Although a Gumbel distribution does not exactly verify the {\small SIR} equation,
the exact solution can be approximated quite well by the Gumbel
distribution by choosing the parameters appropriately. In this section
we will
provide  formulas that relate the parameters of {\small SIR} model
with the parameters of an appropriate Gumbel function. For this end we
will use the proper time method introduced in the previous section.

The Gumbel function is defined here as a time-dependent function with
three parameters
\begin{equation}
G(t)=    a {\rm e}^{ -{\rm e}^{-(t-t_0)/b}   }
\end{equation}
and the derivative gives the Gumbel distribution
\begin{equation}
g(t)= \frac{dG}{dt}= \frac1{b}{\rm e}^{-(t-t_0)/b}G(t). 
\end{equation}
First we will see how well the Gumbel distribution approximates the
rate of infected of the {\small SIR} model, $g(t)\simeq i(t)$.

In
Fig. \ref{fit1} we compare the Gumbel distribution $g (t)$ with the
exact {\small SIR} solution $i (t)$. The parameters of the Gumbel distribution
have been fitted with least-squares method to obtain the optimal
distribution that best describes the exact solution. The parameters of
the {\small SIR} model are $ s_{_0} = 0.99 $, $\beta = 0.3 $, and we use four
values of $ \rho = 1.5, 2, 3$ and  5  ---the same as in Fig. 3. The
fitted parameters of the Gumbel distribution are given in Table 1

The fits of Fig. 4 provide fairly good approximations to the exact {\small SIR}
solutions with the Gumbel distribution. Although the fit is not
perfect, the Gumbel distribution clearly shows similar asymmetry as
the {\small SIR} model. The similarity is greater when the epidemic is
explosive, with a high reproduction number $\rho\sim 3--5$. Where Gumbel fails most in
these fits is in the initial and final stages of the Pandemic.

To investigate the connection between the Gumbel distribution and the
solution of the {\small SIR} equations, it is convenient to express them as 
functions of proper time.
First we substitute $i (t)$ by $g (t)$  in the definition of proper time
\begin{equation}
  d\tau = \beta i dt = \beta g dt = \beta \frac{dG}{dt}dt
\end{equation}
Thus integrating between 0 and $t$ we have
\begin{equation}
  \tau = \int_0^t \beta \frac{dG}{dt}dt=\beta[G(t)-G_0]
\end{equation}
where $G_0=G(0)$. Hence
\begin{equation}
G(t)= G_0+\frac{\tau}{\beta}=  
a {\rm e}^{ -{\rm e}^{-(t-t_0)/b}   }
\end{equation}
Taking the logarithm on both sides we obtain
\begin{equation}
{\rm e}^{-(t-t_0)/b} = \ln\frac{a}{G_0+\tau/\beta}
\end{equation}
from where we obtain the following approximate relation between $t$ and $\tau$
\begin{equation}
t=t_0-b \ln\left(\ln\frac{a}{G_0+\tau/\beta}\right),
\end{equation}
and we can write the Gumbel distribution as a function of $\tau$
\begin{equation} \label{Gumbeltau}
  g=\frac1{b}
  {\rm e}^{-(t-t_0)/b}G =
    \frac1{b}\left(G_0+\frac{\tau}{\beta}\right)\ln\frac{a}{G_0+\tau/\beta}.
\end{equation}

Starting with this expression as a function of proper time, we can
consider several alternative ways of adjusting the Gumbel parameters
from the parameters of the {\small SIR} model, assuming that the proper time is
the same in both models.

\subsection{Proper time fit 1}

The idea of the proper time fit consists in imposing that the maximum
of $g(\tau)$ coincides with the maximum of $i(\tau)$.  In fit 1, we
will also assume that $G_0$ is very small can be neglected, $G_0 \simeq 0$,
i.e., we will not try to
impose any additional condition on the initial value.
This allow us to estimate the parameters $a$
and $b$ easily, but we will not be able to obtain the value of $t_0$,
which will later be adjusted to fit
the temporal peak of the {\small SIR} solution.

We start  by writing Eq. (33) for $G_0=0$
\begin{equation}
  g(\tau)=
 \frac{\tau}{b \beta}\ln\frac{a\beta}{\tau}.
\end{equation}
To find the maximum of this function we compute the derivative
\begin{equation}
  g'(\tau)=
 \frac{1}{b \beta}\ln\frac{a\beta}{\tau}-
 \frac{1}{b \beta}
\end{equation}
from where we find the maximum condition
\begin{equation}
  \ln\frac{a\beta}{\tau}=1.
\end{equation}
Thus the value of the peak position, $\tau_{\rm peak}$, is
\begin{equation} 
  \tau_{\rm peak}= \frac{a\beta}{\rm e}
\end{equation}
and the height of the peak (maximum of $g$) is
\begin{equation} \label{gpeak1}
 g( \tau_{\rm peak})= \frac{1}{\rm e}\frac{a}{b}
\end{equation}
Comparing to the peak values of the {\small SIR} solution,
Eqs. (\ref{taumax},\ref{imax}), and equating the values we obtain
\begin{eqnarray}
\frac{a\beta}{\rm e} &=& \frac{\ln s_{_0}\rho}{\rho}\\
\frac{1}{\rm e}\frac{a}{b} &=&  1- \frac{1+\ln s_{_0}\rho}{\rho}.  
\end{eqnarray}
From here we obtain the values of the parameters $a$ and $b$ of the
Gumbel distribution in terms of the parameters of the {\small SIR} model
\begin{eqnarray}
a &=& \frac{{\rm e}  \ln s_{_0}\rho}{\lambda}\\
\frac{a}{b} &=&  {\rm e} \left(1- \frac{1+\ln s_{_0}\rho}{\rho}  \right)
\end{eqnarray}
We see that the values of $a$ and $b$ can be calculated directly from the
parameters and initial conditions of the {\small SIR} model. The value of $t_0$
would be fitted later to the position of the peak as a
function of time.

Note that if the initial  number of infected individual is very small,
then $s_{_0}$ can be approximated by one in
the logarithm $\ln s_{_0}\rho \simeq \ln\rho$.
Then the parameters $a$ and $b$ 
does not depend appreciably on the precise value of $s_{_0}$,
but only on $rho$ and $\lambda$ by the simple relations
\begin{eqnarray}
a &=& \frac{{\rm e}  \ln\rho}{\lambda}  \label{simplea} \\
\frac{a}{b} &=&  {\rm e} \left(1- \frac{1+\ln\rho}{\rho}  \right)  
\label{simpleab}
\end{eqnarray}
Note also that Eq. (\ref{simpleab}) can be solved numerically to
obtain the reproduction number $\rho$ as a function of $a/b$, and then
Eq. (\ref{simplea}) gives $\lambda$. So the constants $\lambda$ and $\rho$
of the {\small SIR} system can also be computed from the values of the
constants $a$ and $b$ of the Gumbel distribution.

\begin{figure*}[!ht]
\begin{center}
\includegraphics[width=11cm,height=10cm,angle=-0]{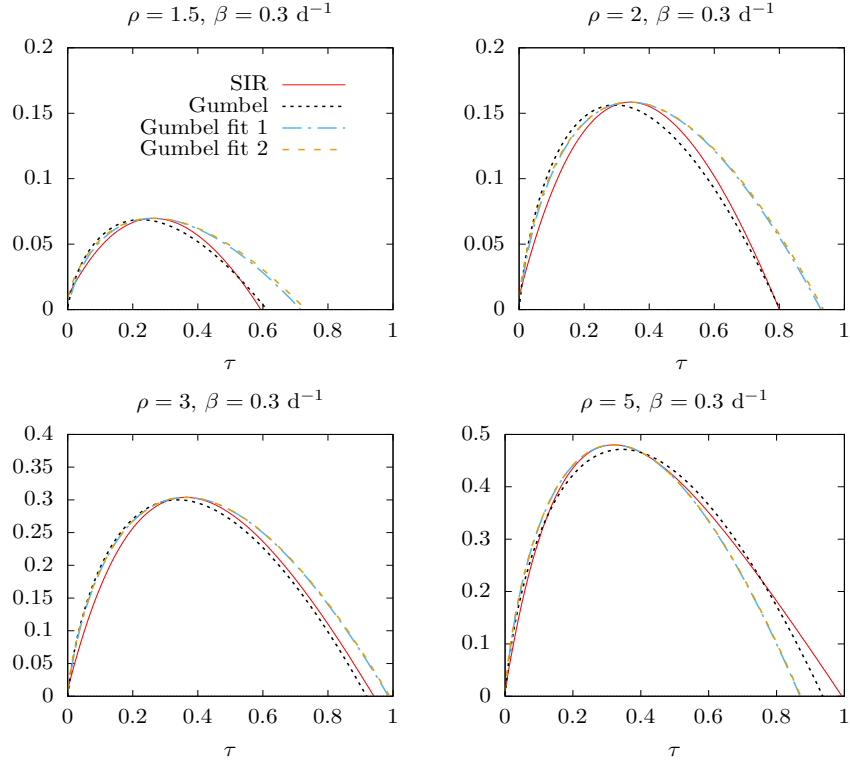}
\caption{\label{fig5} 
  Three Gumbel distributions as a function of the proper time,
compared to the exact
  Solution of the {\small SIR} equations for $s_{_0}=0.99$, $\beta=0.3$ d$^{-1}$,
  and for several values of the reproduction number $\rho$. The
  parameters of the Gumbel distributions correspond to the mean-square fit,
  proper-time fit 1 and proper-time fit 2. In proper-time fit 1, the value of $t_0$ is taken from Tab. 1.}
\end{center}
\end{figure*}

\begin{figure*}[!ht]
\begin{center}
\includegraphics[width=11cm,height=9cm,angle=-0]{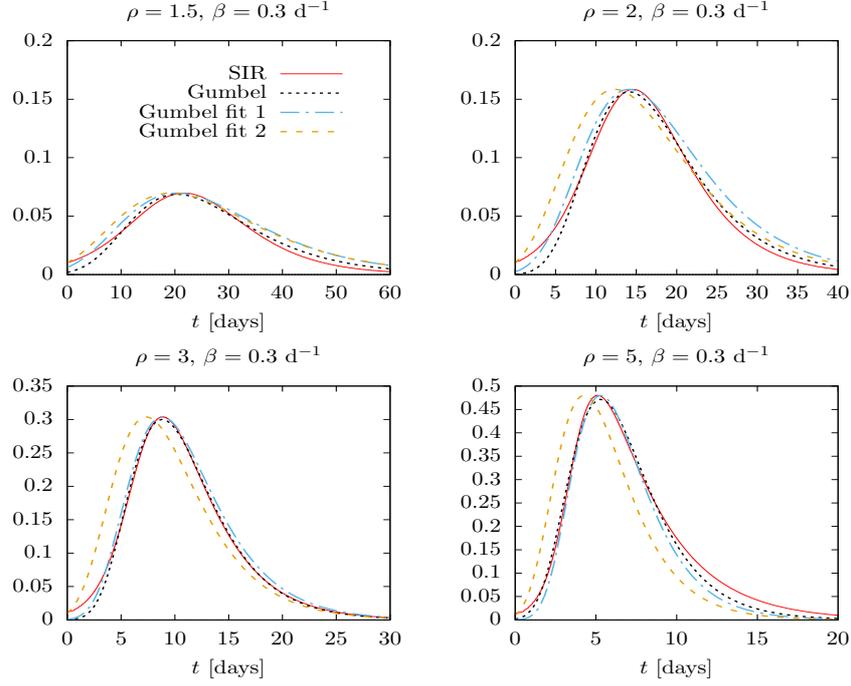}
\caption{\label{fig5} Three Gumbel distributions compared to the exact
  Solution of the {\small SIR} equations for $s_{_0}=0.99$, $\beta=0.3$ d$^{-1}$,
  and for several values of the basic reproduction number $\rho$. The
  parameters of the Gumbel distributions correspond to the mean-square fit,
  proper-time fit 1 and proper-time fit 2. In proper-time fit 1, the
  value of $t_0$ is taken from Tab. 1.}
\end{center}
\end{figure*}

\subsection{Proper time fit 2}

A second fit can be done by fixing also the number of initial infected
$i(0)$, in which case a value for $t_0$ can be theoretically obtained.
We proceed as in previous section, by computing the
maximum of the Gumbel distribution, as a function of proper time,
but in this case we use the exact
expression, Eq. (\ref{Gumbeltau}). The maximum of $g(\tau)$ is now obtained for
\begin{equation}
\ln\frac{a}{G_0+\tau/\beta}=1
\end{equation}
from where the peak (maximum of $g$) is reached at the proper time 
\begin{equation}
\tau_{\rm peak}= \frac{a\beta}{\rm e}-G_0\beta
\end{equation}
and the value of the maximum is
\begin{equation}
g(\tau_{\rm peak})= \frac{1}{\rm e}   \frac{a}{b}
\end{equation}
this is the same value for the peak obtained in Eq. (\ref{gpeak1}).

Comparing to the peak values of the {\small SIR} solution,
Eqs. (\ref{taumax},\ref{imax}), and equating the values we obtain
\begin{eqnarray}
\frac{a\beta}{\rm e} -G_0\beta &=& \frac{\ln s_{_0}\rho}{\rho} \label{apeak2}\\
\frac{1}{\rm e}\frac{a}{b} &=&  1- \frac{1+\ln s_{_0}\rho}{\rho}  \label{abpeak}
\end{eqnarray}
Eq. (\ref{abpeak}) gives the same value for $a/b$ obtained in fit 1. For
$s_{_0}\simeq 1$, this value only depends on the basic reproduction number $\rho$.

A third condition is obtained if we demand $g(0)=i(0)=1-s_{_0}$. Using
Eq. (\ref{Gumbeltau}) for the Gumbel distribution as a function of
proper time, we have
\begin{equation}
g(0)=\frac1{b}\ln\frac{a}{G_0}G_0=1-s_{_0}
\end{equation}
where 
\begin{equation}
G_0= a{\rm e}^{-{\rm e}^{t_0/b}}=a{\rm e}^{-x}
\end{equation}
where we have defined the parameter
\begin{equation}
x={\rm e}^{t_0/b}.
\end{equation}
In terms of the $x$-parameter, the initial value condition can be written as
\begin{equation}
g(0)=\frac{a}{b}x{\rm e}^{-x}.
\end{equation}
The procedure of fit 2 follows the following steps:
\begin{enumerate}
\item Compute $a/b$ from Eq. \ref{abpeak}
\begin{equation}
  \frac{a}{b} =  {\rm e}\left(1- \frac{1+\ln s_{_0}\rho}{\rho}\right).
\end{equation}

\item Write the initial condition $g(0)=i(0)$ as
\begin{equation}
x{\rm e}^{-x}=\frac{b}{a}( 1-s_{_0}) \equiv \epsilon_0
\end{equation}
and solve this equation numerically for $x>0$
\item Once we know the values of $x$ and $a/b$.
compute the value of $a$ using Eq. (\ref{apeak2})
\begin{equation}
a= \frac{\ln s_{_0}\rho}{\lambda({\rm e}^{-1}-{\rm e}^{-x})}
\end{equation} 
\item Compute $b=(b/a) a$, and
  finally
  \item Compute  $t_0$ as 
\begin{equation}  \label{t0}
t_0= b\ln x.
\end{equation}
\end{enumerate}

In Figs. 5 and 6 the solution of the {\small SIR} model is compared with the
Gumbel distributions corresponding to the three fits considered in
this work: the least square fit, and the proper time fits 1 and 2.

The least square fit, with the parameters of Table 1, is
represented with dotted lines as a function of proper time in Fig. 5
and is very close to $ i(\tau) $.  We see in Fig. 5 that the proper
time fit 1 and fit 2 are essentially the same and that their maximum
coincides (by construction) with the maximum of the {\small SIR} solution $ i
(\tau) $.  The parameter $t_0$ in the case of fit 1 is taken from
Table 1, because it cannot be obtained theoretically.
The parameter $t_0$ in fit 2 is computed from Eq. (\ref{t0}) in terms of
the computed values of $b$ and $x$.

For low values of the basic reproduction number, $\rho=1.5--3$, both
fit 1 and fit 2 are wider than $i (\tau)$ in Fig. 5, and they extend
at the end of the epidemic up to a proper time larger than the exact
solution.  When $\rho$ is larger the width of the Gumbel functions of
fits 1 and 2 begin to decrease in relation to the {\small SIR} result,
until their widths become smaller than the width of the {\small SIR}
solution, for $\rho = 5$.

The results of Fig. 5 are translated into the distributions as a
function of physical time in Fig. 6. The Gumbel fits 1 and 2 differ
mainly in the value of $t_0$. The maximum of fit 2 is shifted to the left
with respect to the maximum of fit 1.
This happens because in fit 2 we are demand that the initial
value of the Gumbel distribution be equal to $ i (0)$. As a result
 $g(t)$ is shifted to the left of $i(t)$, because the Gumbel distribution
increases slightly faster than the {\small SIR} solution as a function of
time. By construction, The maximum number of infected coincides
in fits 1 and 2 with the maximum of $i(t)$, but it occurs slightly
earlier in fit 2.

With appropriate parameters, we have seen that the Gumbel distribution
describes the exact solution of the {\small SIR} model quite well. This
justifies that epidemic data can be fitted with
a Gumbel distribution. As a direct application, this would allow
estimating the parameters of the {\small SIR} model in first approximation from
those of Gumbel, using the formulas in this section.
The fact that the Gumbel
distribution is analytical as a function of time allows its parameters
to be easily fitted, unlike the {\small SIR} model, which, although simple,
must be solved numerically,

\section{Asymmetry of the {\small SIR} solutions}

An essential characteristic of the solution $i(t)$ of the {\small SIR}
equations is the evident asymmetry that this function presents with
respect to its maximum value. The asymmetry is more pronounced as the
reproduction number $\rho$ increases. Indeed, in this section we will
define a parameter that uniquely characterizes the asymmetry and we
will see that said asymmetry increases linearly with the basic reproduction number $\rho$

We begin by mentioning an important property of the
$i(t)$ solution concerning the scaling property with respect to time.
It is clearly seen in equation (25) that the physical time as a function
of proper time, $t(\tau)$, is inversely proportional to
$\beta$. Therefore any change in $\beta$ translates into a change in
the time scale. The natural scale of time is thus $\beta t$, that is,
measuring time in units of $1/\beta$. Thus, from the solutions of
the {\small SIR} equations for $\beta=1$, all the others are obtained simply by
re-scaling the time with a factor $1/\beta$.

\begin{figure}[H]
\begin{center}
\includegraphics[width=7cm,height=8.2cm,angle=-0]{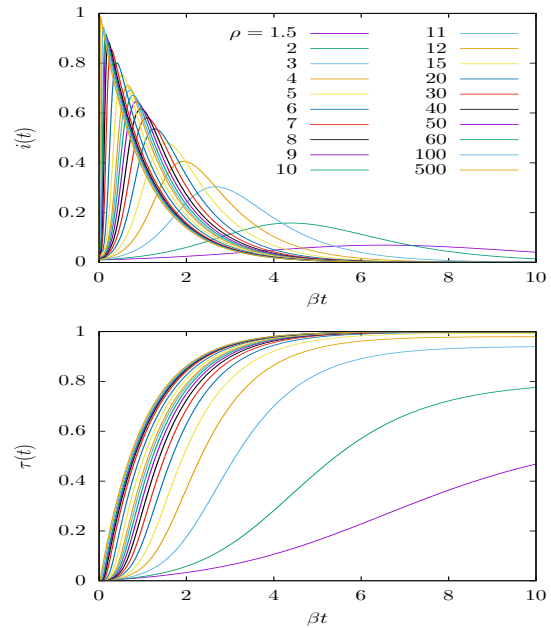}
\caption{\label{asim1} Solutions of the {\small SIR} equations for values of
  the basic reproduction number $\rho$ from 1.5 to 500. The functions $i(t)$
  and $\tau(t)=r(t)$ are Plotted as a function of the normalized time
  $\beta t$.}
\end{center}
\end{figure}

Therefore in Fig. \ref{asim1}, where we represent the solutions of the
{\small SIR} equations for a range of $\rho$ values from 1.5 to 500, and for
$\beta=1$, all the solutions are included if they are represented as a
function of $\beta t$. Note that here we use the initial
condition $s_{_0} = 0.99$. But this does not detract from the generality
of our affirmations, because a change in the initial condition
ultimately translates into a shift of the time origin such that
$i(0)=0.01$.

Figure \ref{asim1} shows that as $\rho$ increases, the epidemic
progresses faster, that is, it ends earlier. As a function of natural time
(or for $\beta=1$ day$^{-1}$), it lasts from about 20 days for
$\rho=1.5$, to less than a week for $\rho>10$, and only a few days if
$\rho > 20$. Moreover
for very large values of $\rho>10$, the epidemic grows
very quickly in a short time interval and then decreases exponentially-like
regardless of the value of $\rho$,
with $i(t) \simeq e^{-\beta t}$. In Fig. \ref{asim1} (top panel)
it is also evident that the asymmetry
of $i(t)$ increases with $\rho$.

\begin{figure*}[!h]
\begin{center}
\includegraphics[width=12cm,height=12cm,angle=-0]{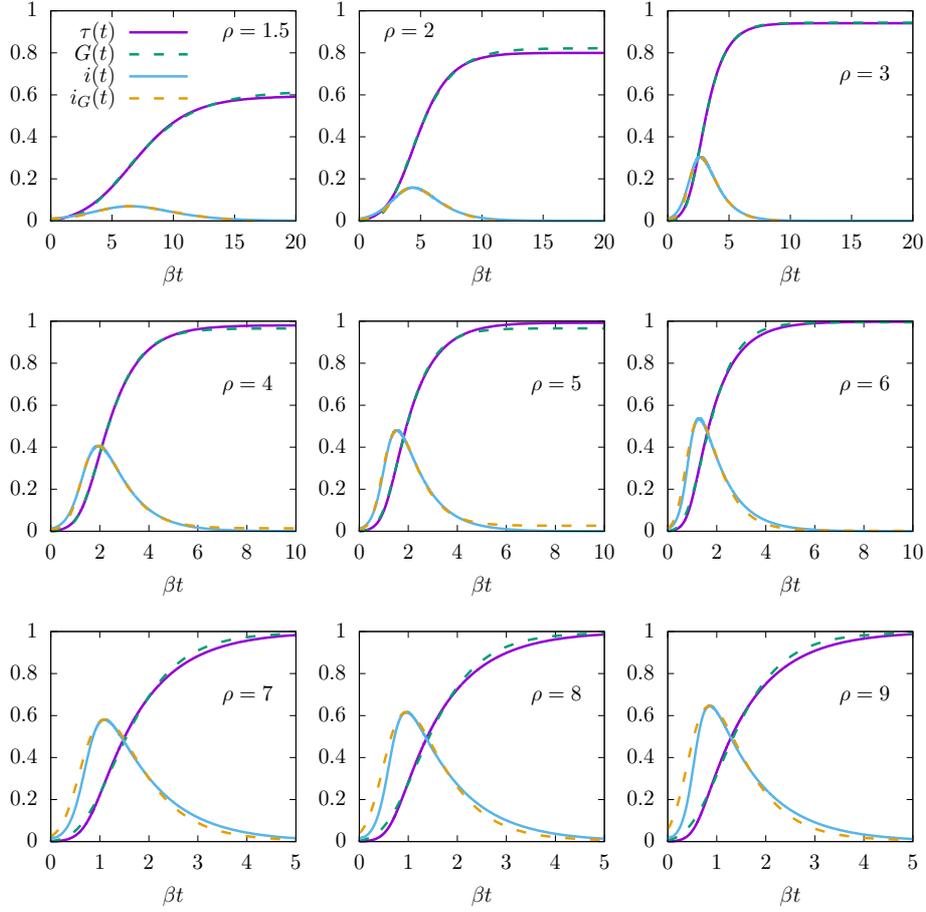}
\caption{\label{fig:asim2} 
  Solutions of the {\small SIR} equations for several values of
  the basic reproduction number $\rho$, and for $s_{_0}=0.99$. The functions $i(t)$
  and $\tau(t)=r(t)$ are plotted as a function of the normalized time
  $\beta t$. These are compared to the approximate solution obtained with a
  Gumbel function $G(t)$ fitted to $\tau(t)$, and with the
  corresponding infected function $i_G(t)= 1- G(t)-s_{_0}\exp(-\rho
  G(t))$ of the extended {\small SIR} model.
}
\end{center}
\end{figure*}

The recovered function, or proper time, $\tau(t)=r(t)$, is displayed
in the bottom panel of Fig. \ref{asim1}. For small values of $\rho$,
it quite resembles a Gumbel function, but for very large values of the
basic reproduction number, $\rho > 10$, it departs from the family of Gumbel
functions and approaches 
the function
$r(t)= 1-\exp(-\beta t)$, as the  limit for $\rho\rightarrow\infty$.

The  behavior of the {\small SIR} solution and the Gumbel
functions can be seen more clearly in Fig. \ref{fig:asim2}. There we plot
the infected and recovered functions as a function of $\beta t$,
for the {\small SIR} model and fitting a Gumbel function, for
$\rho=1.5,2,3
\ldots,9$.
In Fig. \ref{fig:asim2}, the fit is an alternative to the one
made in section 2. There, the Gumbel distribution $g(t)$ was fitted
directly to the infected function $i(t)$. In Fig. \ref{fig:asim2} we first
fit the Gumbel function $G(t)$ to the recovered function
$\tau(t)=r(t)$ by least squares method,
and then we calculate the Gumbel-infected function
using the {\small SIR} Eq. (15)
\begin{equation}
  i_G(t) = 1- G(t) - s_o \exp(-G(t ))
\end{equation}
  for $\beta=1$. The results of fitting $G(t)$ are not exactly the
  same as fitting $g(t)$, because a different function if being fitted
  in each case, but they are very similar.

  We notice that for $\rho=5$, the resulting function $i_G(t)$ does
  not converges to zero for large $t$. This happens because the
  least-squares fit of the Gumbel function does not exactly approach the
  maximum value of $\tau(t)$. To remedy this, starting $\rho=6$, we
  set the value of the $a$ parameter of $G(t)$ so that $G(t)=\tau(t)$
  for large $t$, and then fit the two parameters $b$ and $t_0$.

Fig. \ref{fig:asim2} shows that the Gumbel distribution is very
similar to the {\small SIR} solution, but starting from $\rho=7$, small
differences begin to be seen because the asymmetry of Gumbel and the
{\small SIR} solution begins to diverge from each other.  For $\rho=9$
such a difference is already quite appreciable, and it increases for
larger values of $\rho$.

\begin{figure}[!ht]
\begin{center}
\includegraphics[width=6cm,height=6cm,angle=-0]{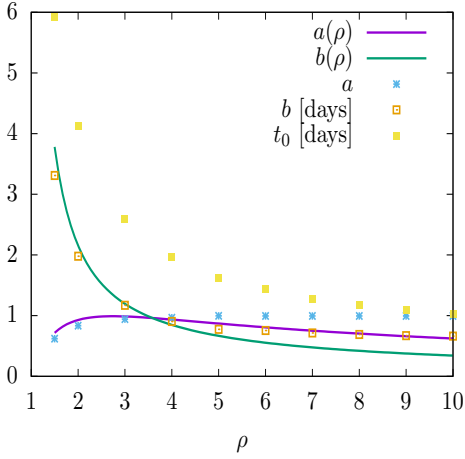}
\caption{\label{asim3} Parameters of the Gumbel function $G(t)$ fitted
  to $\tau(t)$ in Fig. \ref{fig:asim2}. These are compared to the
  functions $a(\rho)$ and $b(\rho)$ from Eqs. (41,42). }
\end{center}
\end{figure}

In Fig. \ref{asim3} we show the values of the Gumbel parameters, $a,
b, t_0$, fitted in Fig. \ref{fig:asim2}, as a function of the
reproduction number $\rho$. We also show the values of $a(\rho)$ and
$b(\rho)$ calculated with Eqs. (41) and (42), which give an analytical
approximation between the Gumbel function and the {\small SIR}
solution. We see that the values of a and b are very close to their
analytic approximations for small $\rho$, and that they start to
differ from $\rho=6$.  This difference is due in part to the fact that
in equations (41) and (42) they were obtained by requiring that the
maxima of $i(t)$ and $g(t)$ coincide. But in Fig.~\ref{fig:asim2} we
have adjusted the parameter $a$ so that the maximum of $r(t)$
coincides with the maximum of $G(t)$ and that alters Eqs. (41) and
(42).

Next we  propose a definition of the asymmetry of the {\small SIR}
solution in terms of the half-widths of the distribution $i(t)$ to the
left and to the right of the maximum $i_{peak}=i(\tau_{peak})$.
First we define $\tau_1$ and $\tau_2$ as the values 
of the proper time such that $i(\tau_1)= i(\tau_2)= i_{peak}/2$ and 
$\tau_1<\tau_{peak}<\tau_2$.
Specifically, $\tau_1$ and $\tau_2$ are the two solutions of the equation
\begin{equation} \label{ipeak2}
1-\tau-s_{_0}{\rm e}^{-\rho\tau}= \frac{i_{peak}}{2}
\end{equation}
where $\tau_{peak}$ and $i_{peak}$ corresponds to the position and
value of the maximum of $i(\tau)$, given by Eqs. (19, 20).
Therefore $\tau_1$ and $\tau_2$ are the proper-time values at which 
$i(\tau)$ reaches half of its maximum value. 
Two half-widths are defined in term of the physical time as
\begin{eqnarray}
\Gamma_1 & = & \beta(t(\tau_{peak})-t(\tau_1))\\
\Gamma_2 & = & \beta(t(\tau_2) - t(\tau_{peak}))
\end{eqnarray}
In view of the previous numerical results we see that $\Gamma_2 >
\Gamma_1$ for $\rho>1$, that is, the {\small SIR} solution is wider to the
right than to the left of the maximum.  The asymmetry is then defined
as the quotient
\begin{equation}
  {\cal A}(\rho)= \frac{\Gamma_2}{\Gamma_1} \ge 1.
\end{equation}
This function increases with $\rho$.  This asymmetry is defined as a
ratio that does not depend on $\beta$ nor on the global normalization
of $i(t)$. Therefore, it is a suitable parameter to express the
asymmetry of the theoretical distribution, as well as that of the
experimental data of deaths described in the next section.

\begin{figure}[!ht]
\begin{center}
\includegraphics[width=6cm,height=6cm,angle=-0]{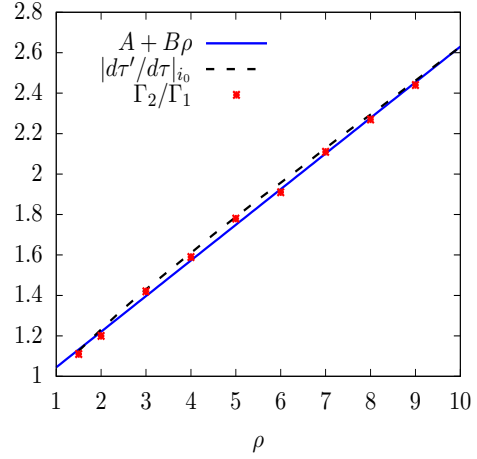}
\caption{\label{asim4} Asymmetry of the {\small SIR} solutions as a function of
  $\rho$.  The points are the values of the asymmetry computed
  numerically for discrete values of $\rho=1.5, 2, 3,\ldots,9$. The
  solid line is a fitted right line $A+B\rho$. The dashed line is the
  Jacobian $|d\tau'/d\tau|_{i_{_0}}$ evaluated for $i_{_0}=0.84 i_{peak}$.
}
\end{center}
\end{figure}

To compute the asymmetry of the {\small SIR} solution.
for a $\rho$ value, we fist solve the
transcendental equation (\ref{ipeak2}) numerically, and obtain the two solutions
$\tau_1 < \tau_2$. The half widths are then computed as the
integrals
\begin{eqnarray}
  \Gamma_1
  &=& \int_{\tau_1}^{\tau_{peak}}\frac{d\tau}{1-\tau-s_{_0}{\rm e}^{-\rho \tau}}
    \\
 \Gamma_2
    &=& \int_{\tau_{peak}}^{\tau_2}\frac{d\tau}{1-\tau-s_{_0}{\rm e}^{-\rho \tau}}  
\end{eqnarray}
These integrals are computed numerically and the corresponding
asymmetry is plotted in Fig. \ref{asim4} for $\rho=1.5,2,3,\ldots,9$,
and for $s_o=0.99$.  We see that, as a function of $\rho$, the
asymmetry of the {\small SIR} solutions is quite approximately a
linear function of $\rho$, which is very well fitted by the
parametrization ${\cal A}= A + B\rho$, with $A=0.868$ and $B=0.176$,
as seen in Fig. \ref{asim4}.

To better understand why this quasi-linear dependence on asymmetry
occurs, we proceed as follows. A rigorous proof is not possible to our
understanding because transcendental equations are involved, and
because the linearity is only approximate, but its origin can be
roughly understood.

First, in the interval $[\tau_1,\tau_{peak}]$ the function
$i(\tau)=1-\tau-s_{_0}{\rm e}^{-\rho\tau}$
is increasing and the change of variable $\tau\rightarrow i$ can be made
inside the integral $\Gamma_1$,
with Jacobian $di/d\tau= \rho s_{_0}{\rm e}^{-\rho\tau}-1 >0$.
We obtain
\begin{equation}
  \Gamma_1 = \int_{i_{peak}/2}^{i_{peak}}\frac{d\tau}{di}
\frac{di}{i}
\end{equation}
Second, in the interval $[\tau_{peak},\tau_2]$ the function
$i(\tau')=1-\tau'-s_{_0}{\rm e}^{-\rho\tau'}$
is decreasing and the change of variable $ i\rightarrow \tau'$ can be made
inside the integral (65),
with Jacobian $di/d\tau'= \rho s_{_0}{\rm e}^{-\rho\tau'}-1 <0$.
We obtain
\begin{equation}
  \Gamma_1
=
  \int_{\tau_2}^{\tau_{peak}}\frac{d\tau}{di}\frac{di}{d\tau'}
  \frac{d\tau'}{i(\tau')}
=
\int_{\tau_2}^{\tau_{peak}}\frac{d\tau}{d\tau'}
\frac{d\tau'}{i(\tau')}
\end{equation}
with
\begin{equation}
\frac{d\tau}{d\tau'}=
\frac{d\tau}{di}
\frac{di}{d\tau'}= 
\frac{ \rho s_{_0}{\rm e}^{-\rho\tau'}-1 } {\rho s_{_0}{\rm e}^{-\rho\tau}-1}<0\, .
\end{equation}
This Jacobian corresponds to the change of variable $\tau\rightarrow\tau'$,
inside $\Gamma_1$,
where $\tau <\tau'$ are the two solutions of the transcendental equation
$i=1-\tau-s_{_0}{\rm e}^{-\rho\tau}$,
for a fixed value of $i \in [i_{peak}/2,i_{peak}]$.

Now, using the mean value theorem, we can factor the Jacobian out of
the second integral of Eq. (66),
evaluated at some intermediate value of $i=i_{_0}$ between
$i_{peak/2}$ and $i_{peak}$, obtaining
\begin{equation}
  \Gamma_1
 =\left|\frac{d\tau}{d\tau'}\right|_{i_{_0}} 
\int_{\tau_{peak}}^{\tau_2}
\frac{d\tau'}{i(\tau')}
 =\left|\frac{d\tau}{d\tau'}\right|_{i_{_0}} 
\Gamma_2
\end{equation}
Computing numerically both sides of (68) for different values of
$i_{_0}$ we found that the best
value is $i_{_0}=0.84$, for which we plot the Jacobian as a function of
$\rho$ in Fig. \ref{asim4}. We see that quite a straight line is
obtained that coincides with the calculated value of the asymmetry.

The quasi-linear behavior of the asymmetry of the {\small SIR} solutions as a
function of $\rho$ allows us to obtain the value of $\rho$ from experimental
data, after estimating the asymmetry empirically.

\begin{table*}
\begin{center}  
\begin{tabular}{ccccccc}  \hline
          &$G(t)$  & $g(t)$ & {\small SIR}  &                &             &$g(t)$ fit to {\small SIR}
  \\
  Country &$b$ [d] &$b$ [d] &$\rho$&$\beta$ [d$^{1}$]&$\beta^{-1}$ [d] & $b$ [d]
  \\  \hline
  Spain   & 13.4   &12.6    & 7    & 0.05           & 20           & 12 \\
  UK      & 17.7   &16.15   & 7    & 0.038          & 26           &16.2\\
  France  & 13.35  &11.6    & 6    & 0.06           & 16.7         & 10.8\\
  Italy   & 18.4   &17.1    & 5.5  & 0.045          & 22.2         & 15.2\\
  Germany & 15.4   &14.6    & 5.5  & 0.05           & 20           & 13.6\\
  Canada  & 18.01  &18.1    & 5.1  & 0.041          & 24.4         & 16.9\\
  Belgium & 12.8   &11.65   & 3    & 0.1            & 10           & 11.27\\
  Switzerland&11.7   &11.79   & 4    & 0.085        & 11.8         & 10.16\\
  Sweden  & 26.7   & 25.66  & 7    & 0.03           & 33           & 18.8\\
  USA     & 21.1 [0:150] & 19.6  & 8    & 0.028     & 35.7         & 17.7\\
          &        & 47.4 (2nd wave)&  &            &              &     \\
  Brazil  & 60.6 [0:250] & 35.3 &  &                &              &     \\
          &              & 37 (2nd wave) & &        &              &     \\
 India    & 60           & 66.4          & &        &              &\\ 
          &              & 32 (2nd wave) & &        &              &  
  \\ \hline
\end{tabular}
\end{center}
\caption{Parameters of the Gumbel and {\small SIR} models fitted for the different countries considered in this work.}
\end{table*}

\begin{figure*}
\begin{center}
\includegraphics[height=11cm,angle=-0]{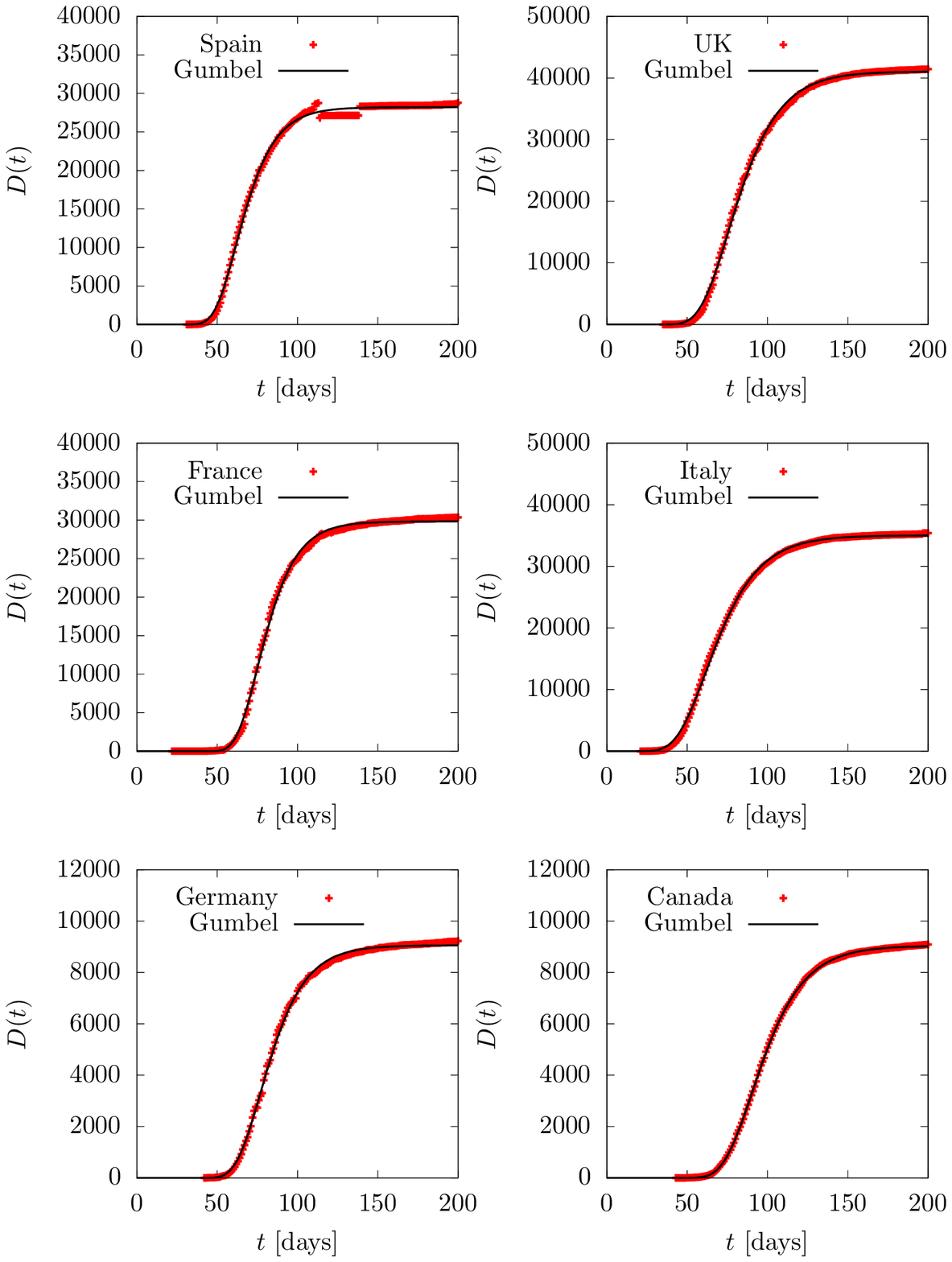}
\caption{\label{fig7} 
  Total deaths $D(t)$ during the first wave of the {\small COVID}-19 pandemic for several countries,
  compared to the Gumbel function $G(t)$.
  The parameter $b$ of the Gumbel function fitted to the data is given in Table 2. Data are form Ref.\cite{world}. 
}
\end{center}
\end{figure*}

\begin{figure*}[!ht]
\begin{center}
\includegraphics[height=11cm,angle=-0]{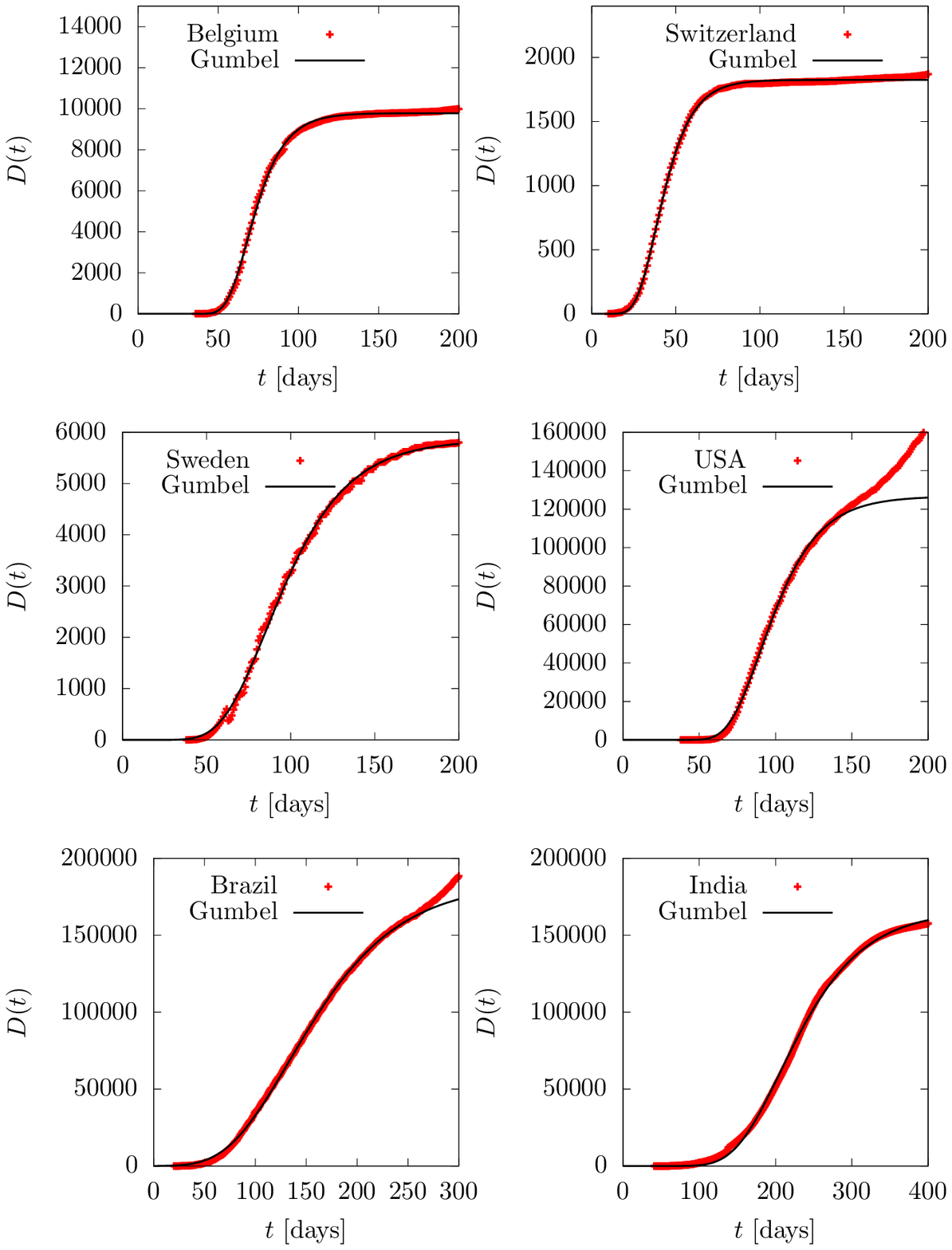}
\caption{\label{fig10} 
  Total deaths $D(t)$ during the first wave of the {\small COVID}-19 pandemic for several countries,
  compared to the Gumbel function $G(t)$.
  The parameter $b$ of the Gumbel function fitted to the data is given in Table 2. Data are form Ref.\cite{world}. 
}
\end{center}
\end{figure*}

\section{Results}

In this section we present results comparing with the data of total
deaths $ D (t) $ and daily deaths $ \Delta D (t) $ as a function of
time for the first wave of the Covid-19 pandemic.  These data are
supplied by government ministries in different countries and are
available from various sources. 
 The data can vary in different sources, as it can often be found
averaged or smoothed, or in many cases the data can be later revised
and updated differently at different sites.

To study the daily deaths we are assuming that
a certain constant portion of removals ends in death, that is,
the death rate verifies an equation similar to the removal rate
with a different constant
\begin{equation}
  \frac{dD}{dt}  =   \alpha I. 
  \end{equation}
this means that the daily deaths,
$\Delta D(t) \simeq   \frac{dD}{dt}$, 
 are described, except normalization
constant, by the solution $i(t)$ of the SIR equations and we can use the
formalism of the previous sections.

One of our purposes of this work is to check if the {\small SIR} model
is capable of describing the data, in which case it can be assumed
that the hypotheses of this very simple model are justified; in
particular if averaged values can be adopted for the two basic
constants of the model: the reproduction number $\rho$ and the
recovery rate $\beta $. This would indicate the universal validity of
the {\small SIR} model. Comparing the parameters of the model between
different countries will tell us the degree of universality of these
parameters when passing from one country to another.

For the present study we have selected the countries where the first
wave of the  {\small COVID}-19 pandemic is visually similar to the solution
of the {\small SIR} equations. The data that we have studied come from
Ref. \cite{world}. But they can also be found in other sources, such
as the worldometer website~\cite{worldometer}.
Surprisingly, after an inspection of the death data for each country,
it turns out that the solutions of the
 {\small SIR} equations can only be applied
in a handful of countries, which can
be counted with the fingers of both hands. These countries are:
Spain, France, Italy,
UK, Germany, Canada, Belgium, Switzerland, Sweden and USA (we left
aside the curious case of China where the epidemic ended mysteriously
and prematurely). Here we only consider the first wave of the pandemic,
because the
subsequent ones are much more irregular and require a separate study.
Note that we only analyze the data on daily deaths and cumulative
deaths. The homogeneity of the reported number of infected individuals
is questioned because it is
proportional to the number of tests performed, and the
experimental error of the tests is not given.

To begin with, in Figs. \ref{fig7} and \ref{fig10} we show the
accumulated deaths as a function of time in the ten countries
mentioned with the addition of Brazil and India, as they are the
top countries in number of infections and perhaps also in number of
deaths.  Time is measured in days.
Data are from Ref.\cite{world} and day one is 2020 02
01. A Gumbel function has been fitted for each country. The only
 tabulated parameter in the second column of Table 2 is
 the value of the $b$ parameter in days.
 Note that the value of $a$ in the Gumbel function
 is simply a normalization constant, and the value of $t_0$ is a shift in
 time. Thus what really characterizes the dynamics
 is the value of the $b$,
which is related to the duration of the epidemic.  Note that the
USA data have been fitted until day 150, when the second wave
starts to appear. In the case of Brazil and India the fit is performed
until day 250.  Note that
Spain, France and Belgium and Switzerland have similar $b$-values in the range
$b\sim 12$--13 days. Italy, UK
and Canada are well fitted with $b\sim18$ days. Germany in in between with $b\sim 15$, and 
Sweden is the European country with the highest value $b\sim 27$ days.

\begin{figure*}[!ht]
\begin{center}
\includegraphics[width=11cm,height=13cm,angle=-0]{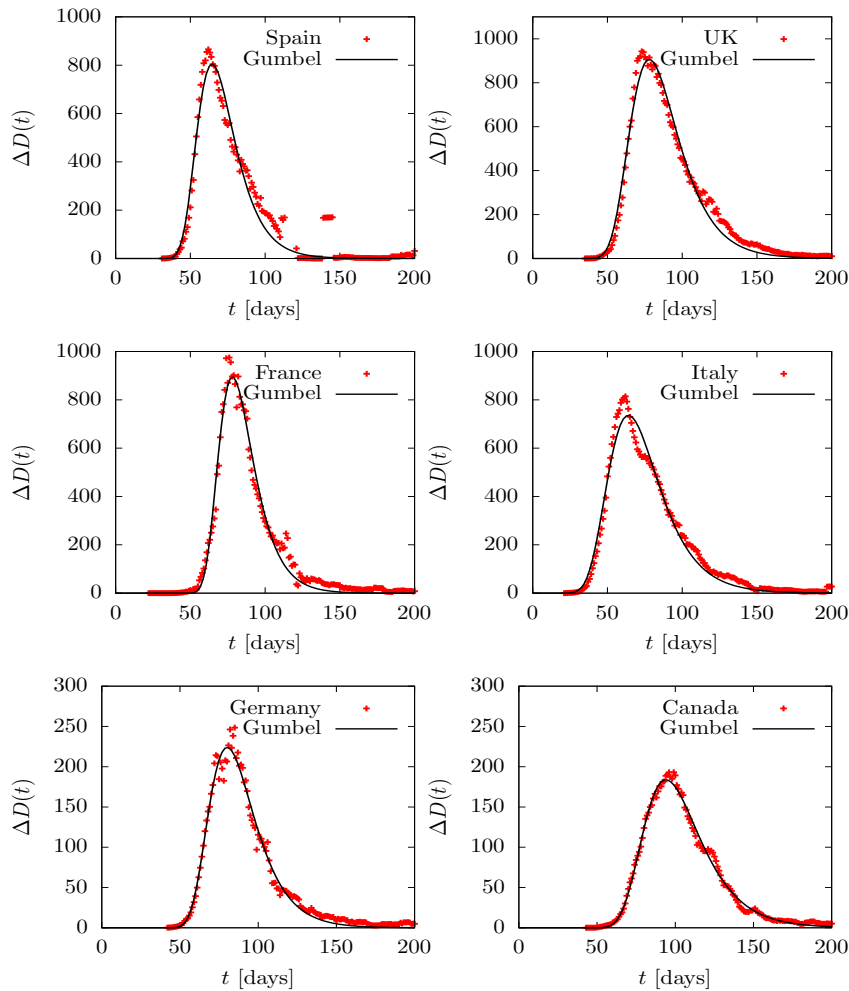}
\caption{\label{fig8} 
  Total deaths $D(t)$ during the first wave of the {\small COVID}-19 pandemic for several countries compared to the Gumbel function $G(t)$.
  The parameter $b$ of the Gumbel function fitted to the data is given in Table 2. Data are form Ref.\cite{world}. 
}
\end{center}
\end{figure*}

\begin{figure*}[!ht]
\begin{center}
\includegraphics[width=11cm,height=13cm,angle=-0]{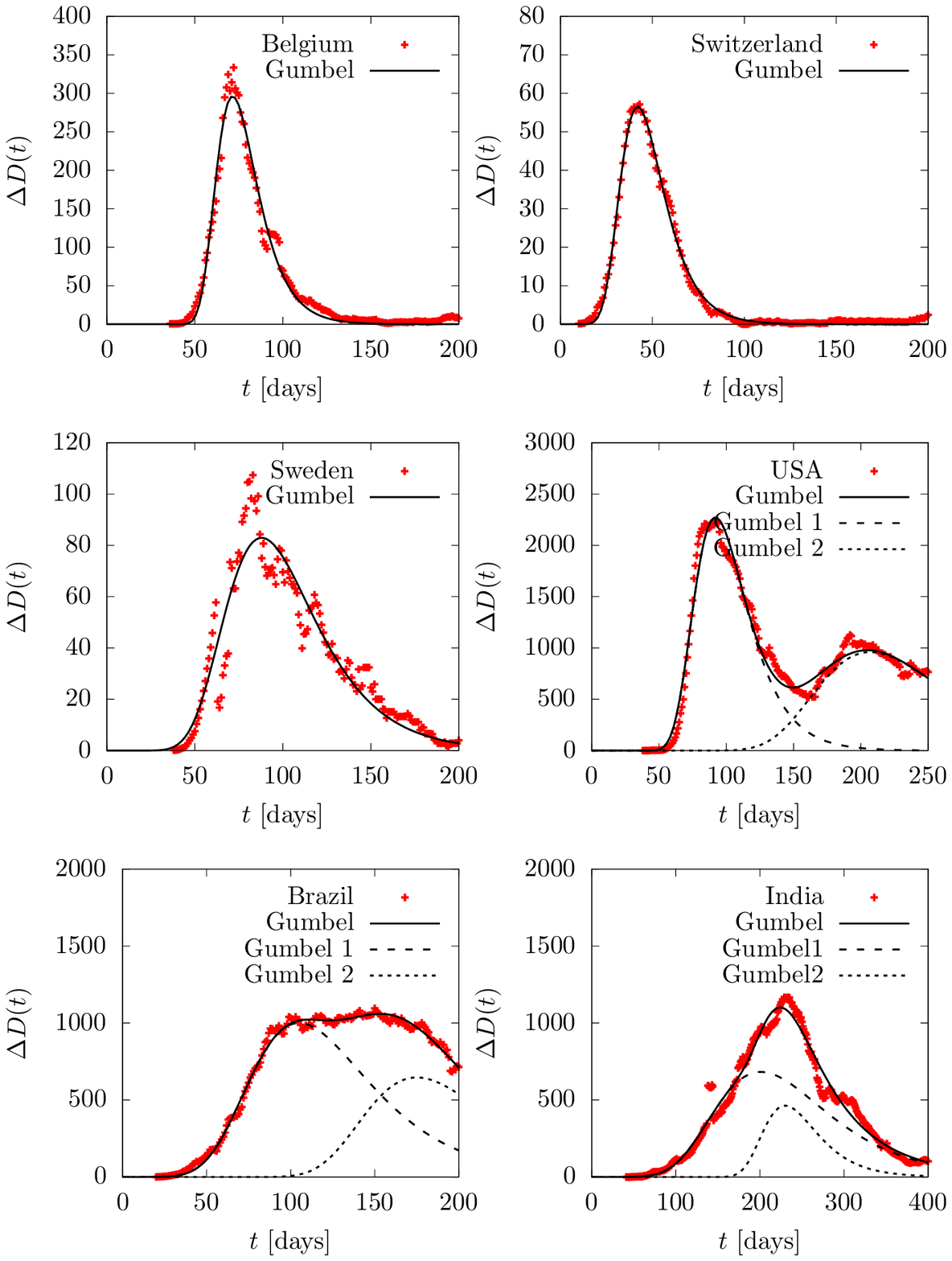}
\caption{\label{fig11} 
  Total deaths $D(t)$ during the first wave of the {\small COVID}-19 pandemic for several countries compared to the Gumbel function $G(t)$.
  The parameter $b$ of the Gumbel function fitted to the data is given in Table 2. Data are form Ref.\cite{world}. 
}
\end{center}
\end{figure*}

Cumulative deaths, $D(t)$, are fairly smooth distributions and very similar
among different countries
because we are dealing with sums ---or integrals on the continuous
limit. More detailed information is obtained by describing the daily
death data $\Delta D(t)$, shown in Figs. \ref{fig8} and \ref{fig11}.
Although these data show more fluctuations, they can be fitted well
with the Gumbel $g(t)$ distribution, although the fit parameters
differ somewhat from those obtained by fitting the Gumbel function
$G(t)$, since different functions and data are being involved.  the
fitted parameter $b$ is in the third column of Table 2. Again we only
tabulate the parameter $b$, because $a$ and $t_0$ give simply the
relative height and the position of the peaks.  in the case of the
USA, Brazil and India, we fit two Gumbel distributions, since it is
apparent that there are at least two overlapping waves. In these
cases, in Table 2 we tabulate the two values $b_1$ and $b_2$ of the
two adjusted waves.

\begin{figure*}[!ht]
\begin{center}
\includegraphics[width=11cm,height=13cm,angle=-0]{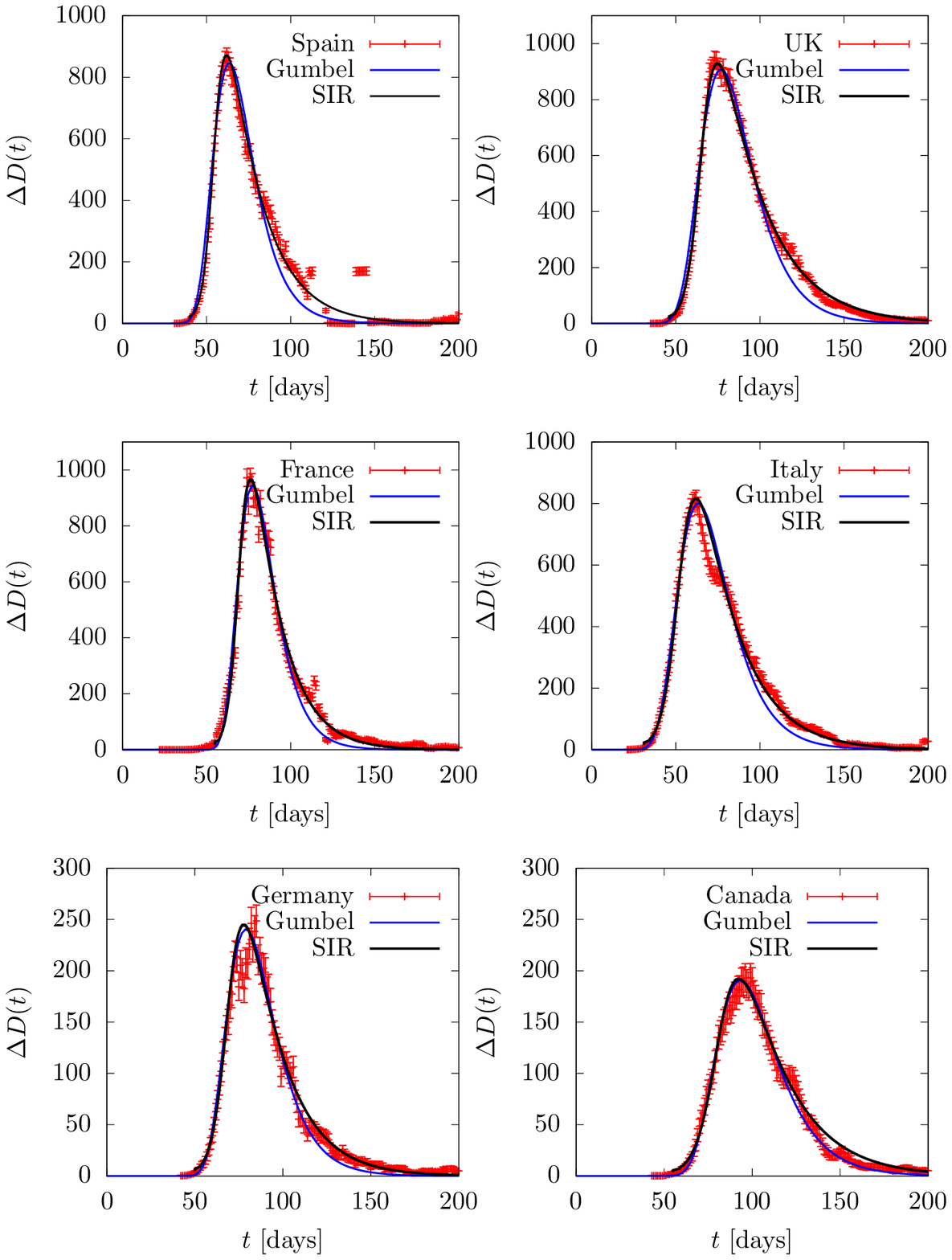}
\caption{\label{fig9}  Total deaths $D(t)$ during the first wave of the {\small COVID}-19 pandemic for several countries compared to the Gumbel function $G(t)$.  The parameter $b$ of the Gumbel function fitted to  the data is given in Table 2. Data are form Ref.\cite{world}.  }
\end{center}
\end{figure*}

\begin{figure*}[!ht]
\begin{center}
\includegraphics[width=10cm,height=9cm,angle=-0]{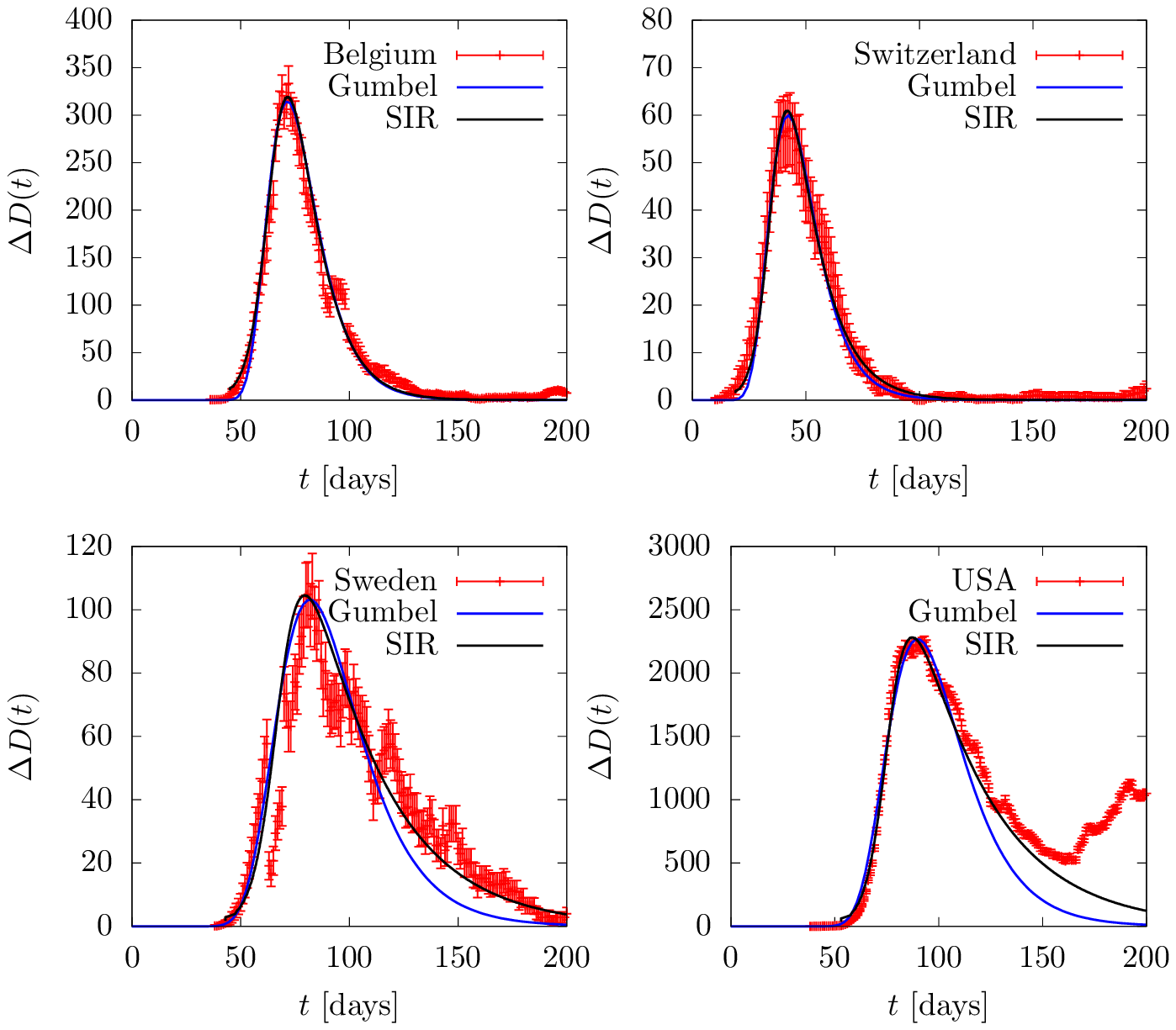}
\caption{\label{fig12} Total deaths $D(t)$ in the first wave of
  coronavirus pandemic for several countries, compared to the Gumbel
  function $G(t)$.  The parameter $b$ of the Gumbel function fitted to
  the data is given in Table 2. Data are form Ref.\cite{world}.  }
\end{center}
\end{figure*}

We see that all countries fit one or two Gumbel distributions quite
well, so this function is an optimal candidate to quantitatively
describe an epidemic of these characteristics with only one parameter,
$b$, plus the normalization and position of the peak. The fact that
the nine countries considered with an isolated first wave (Spain,
France, Italy, UK, Germany, Canada, Belgium, Switzerland and Sweden)
only require a time-independent parameter is remarkable. This does not
happen in the following waves or in other countries, where the data
show different behavior
with large overlaps and stochastic fluctuations.

Since Gumbel provides a good analytical approximation to the {\small SIR} model
solution, it is natural to wonder if the exact {\small SIR}
solution would give an
even better description of the data. So in Figs. \ref{fig9} and
\ref{fig12} we compare data with exact {\small SIR} solutions given by the
equations of the previous sections. We know from the last section that
there is a linear relationship between
the asymmetry of data and the basic reproduction number,
$\rho$. This has allowed us to obtain an approximate value
of $\rho$ and then we have fitted the value of
$\beta$ and a normalization factor to the width and height of the
data, we have added a shift  in time to get the position of the peak.
The parameters $\rho$, and $\beta$ are given in Table 2.  
In Fig. \ref{fig12} we have fitted the US data only to where the
first peak is clearly seen. For this reason the other two countries, Brazil
and India have not been fitted.

In Figs. 15, 16,
we also plot the results of the  Gumbel
distribution, but this time the parameters have not been fitted to the
data, but to the respective {\small SIR} solution.
Note that, from Eqs. (43) and (44) 
\begin{equation} \label{betab}
 \beta b=  \frac{\ln \rho}{\rho-1-\ln\rho}
\end{equation}
From inspections of the numbers given  Table 2, columns 4, 5 and 7,
we see that this equation is approximately verified. For instance, for
$\rho=7$ (Spain and UK),
the right-hand side of Eq. (\ref{betab}) gives 0.48, and
roughly $1/\beta\simeq 2b$.
For $\rho=3$ the equation gives $1/\beta\simeq b / 1.2$.

We see that in general the fit of the Gumbel distribution to the {\small SIR}
solutions in these countries is good, although it begins to fail, for
high values of $\rho$, in the tail part, as we have already seen in the
previous section

From Figs.~\ref{fig9} and \ref{fig12} we conclude that the data of
the countries considered are globally well described with an exact {\small SIR}
solution, without the need for any time dependence of the
parameters. Again we underline that this only happens in the first
wave of the countries that we have considered here and not in the
other countries or in the remaining waves. The cause of this requires
a detailed study of the epidemiological causes that is beyond the
scope of this paper.

\begin{figure}[!ht]
\begin{center}
\includegraphics[width= 7.5cm]{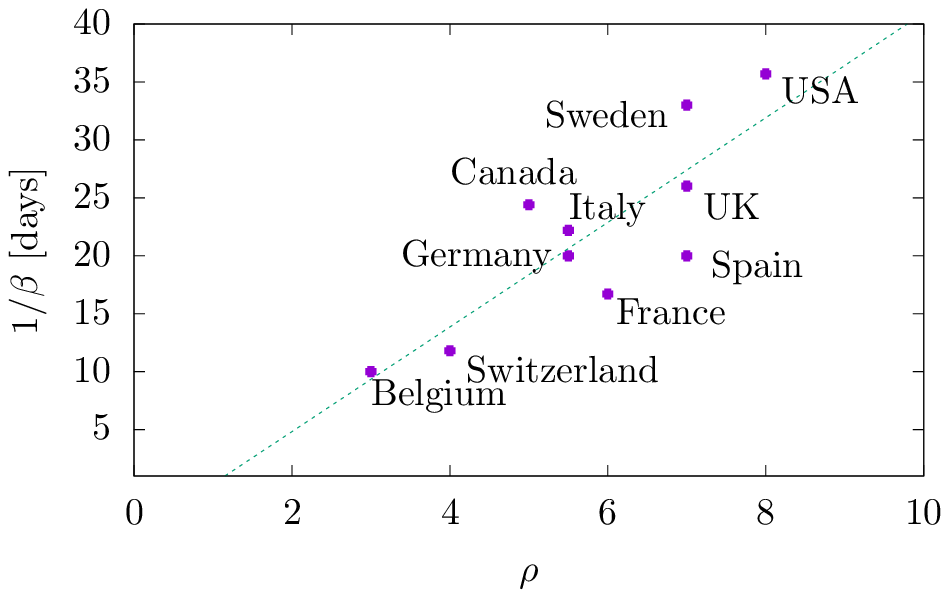}
\caption{
  \label{paises}
  Locations of the different countries studied in the plane of {\small SIR}
  parameters $(\rho, 1/beta)$.
}
\end{center}
\end{figure}

Finally, in Fig.~\ref{paises} we plot the values of the
parameters of the {\small SIR} solution in the $(\rho, 1/\beta)$ plane. If we
fit a right line to these data,
we see that the general trend seems to be for $1/\beta$ to
increase with $\rho$. If we exclude the 'outsider' countries, Belgium and
Switzerland (which have small values of $\rho= 3$ or 4) and Sweden and
USA (with large values of $\rho=7$, 8), we are left with six countries
with similar parameters, in the central zone, around $\rho =6$ and
$1/\beta= 20$ d ---Spain, France, Italy, UK, Canada and Germany. In
these countries a relationship between $\beta$ and $\rho$ is not
found.  On the other hand, the fact that $\beta$ is similar in these six
countries indicates that the probability of recovery is similar in all
of them.
The recovery probability $\beta$ for an individual should in principle be
independent of location. But if we consider that there could be
important effects due to medical treatments and hospital capacity to
treat severe cases in different countries, this could explain the
differences between the $\beta$ values.

Note that in most countries the value of $\rho$ exceeds five,
indicating an explosive increase in the first stage of the
epidemic in each country. An important consequence of the {\small SIR}
equations for these high values of $\rho$
is that the total number of people infected during the
epidemic reaches almost 100\%. Indeed, this is mathematically
given by the value of the proper time $\tau$ at the end of the epidemic,
which is the solution of Eq. (16). Numerically it is easy to verify
that this end value is practically one, for $\rho > 5$. This can also be seen
in the lower panel of Fig.~7, where the value of $\tau(t)=r(t)$ 
 for large $t$
is practically one. As a consequence, our results indicate that the
data from these countries where $\rho>5$ are compatible with an epidemic
where practically all initially susceptible individuals were infected,
according to the {\small SIR} model.

At this point we can already link with the question raised in the
introduction of this work, which is whether the lock-downs and other restrictions over the population had any
effect in reducing mortality. According to the meta-data study of
Ref.\cite{Her22}, the NPI had practically no effect on
mortality. This seems to be corroborated in our study for three
reasons: (i) that the data are compatible with {\small SIR} solutions with high
value ($R_0=\rho\simeq 6$) of the basic reproduction number, which implies
that all the susceptible individuals were infected; (ii) that the {\small SIR}
parameters do not depend on time, but if there were some NPI effects the
parameters should be time dependent and the
epidemiological curves should differ
from the {\small SIR} solutions; (iii) there does not seem to be a relationship
between the intensity of the lock-down measures and the basic
reproduction number. For example, Spain, which had very harsh
restrictions, is fitted with the same reproduction number as Sweden,
which practically did not have, and is greater than Italy, where the
measures were introduced a week earlier.

\section{Conclusions} 

In this work we have solved the differential equations of the {\small SIR}
model in a parametric way using the proper time as a parameter,
defined as the relative number of recovered individuals $\tau=r(t)$. As a
function of $\tau$, the {\small SIR} solution is analytical, which allows us to
study some of its properties, such as, for example, calculating the
maximum number of infected $i_{\rm peak}$ and the asymptotic number of recovered
at the end of the epidemic $r(\infty)$.

Secondly, we have studied the possibility of approximating the {\small SIR}
solutions using Gumbel distributions $g(t)$, because this family of
functions presents a similar asymmetry as the {\small SIR} solutions, and 
only depends on one parameter, plus the normalization and the
position. We have proposed various methods of fitting Gumbel
distributions to exact {\small SIR} solutions. In particular, using the proper
time, we have found simple relationships between the Gumbel parameters
and the {\small SIR} model parameters.

Third, we have discussed the scaling properties
of the exact {\small SIR} solutions when
plotted as a function of $\beta t$, where $\beta$ is the probability
of removal per unit of time. Next we have defined an asymmetry
parameter, as the ratio between the right and left half-widths at
half-height, of the {\small SIR} solutions. We have shown numerically that the
asymmetry ${\cal A}(\rho)$ grows almost linearly with the reproduction
number $\rho$ and that it is independent of $\beta$. The asymmetry
uniquely characterizes the value of $\rho$ and vice-versa.  Therefore,
a measure of the asymmetry of some data of $i(t)$ at the middle of the
height, allows to extract the value of $\rho$.

Finally, we have applied the {\small SIR} model and the Gumbel distribution to
study the daily death-data in the first wave of  the {\small COVID}-19 pandemic
in a dozen countries. The countries have been selected because they
are the only ones that present a peak that closely resembles a
{\small SIR} solution. The data from Spain, France, Italy, UK, Canada, Germany,
Belgium, Switzerland and Sweden can be fitted
quite well with a {\small SIR} solution and
also with a Gumbel function. Except for Belgium and Switzerland, data
in the rest of the countries are compatible a reproduction number
$\rho>5$. This seem to indicate that in practically all
susceptible individuals were infected and eventually recovered.  This raises
questions about the effectiveness of non-pharmaceutical interventions,
such as lock-downs, in many
countries.

\section{Acknowledgments}
The author thank Dr. Nico Orce for critical reading of the manuscript.
This work has been supported by the Spanish Agencia Estatal de investigacion
(D.O.I. 10.13039/501100011033, 
Grant No. PID2020-114767GB-I00), 
the Junta de Andalucia (Grant No. FQM-225).

%%%%%%%%%%%%%%%%%%%%%%%%%%%%%%%%%%%%%%%%%%%%%%%%%%%%%%%%%%%%%%%%%%%


\begin{thebibliography}{99}



\bibitem{Hui20}  D. S. Hui, E. Azhar, T. A. Madani {\it et al.}., The continuing 2019-nCoV epidemic threat of novel
coronaviruses to global health — The latest 2019 novel coronavirus outbreak in Wuhan, China. Int. J. Infect. Dis. {\bf 91} (2020) 264. 


\bibitem{Tan20} B. Tang, X. Wang, Q. Li, N.L. Bragazzi, S. Tang, Y. Xiao, and J. Wu, 
Estimation of the transmission risk of the 2019-nCoV and its implication for public health interventions, 
Journal of Clinical Medicine {\bf 9} (2) (2020) 462.


\bibitem{Wu20} J. T. Wu, K. Leung, and G. M. Leung, 
Nowcasting and forecasting the potential domestic and international spread of the 2019-nCoV outbreak originating in Wuhan, China: a modelling study, The Lancet {\bf 395} (10225) (2020) 689.


\bibitem{Kra20} M. U. G. Kraemer, C. H. Yang, B. Gutierrez, C. H. Wu, B. Klein, D. M. Pigott, and J. S. Brownstein, 
The effect of human mobility and control measures on the {\small COVID}-19 epidemic in China. Science {\bf 368} (6490) (2020) 493.


\bibitem{Ham06} W. H. Hamer,  The Milroy Lectures on Epidemic disease in England -- the evidence of variability and of persistency of type. 
The Lancet {\bf 167} (4305) (1906) 569. 

\bibitem{Ros08} R. Ross,  Report on the Prevention of Malaria in Mauritius. London: Waterlow and Sons (1908). 

\bibitem{Ros16} R. Ross, An application of the theory of probabilities to the study of a priori pathometry. -- Part I. 
Proc. R. Soc. Lond. A {\bf 92} (1916) 204. 

\bibitem{Ros17} R. Ross and H. P. Hudson, 
An application of the theory of probabilities to the study of a priori pathometry.--Part III.
Proc. Roy. Soc. A {\bf 93} (1917) 225.


\bibitem{Ker27} W. O. Kermack and A. G. McKendrick.  
A contribution to   the mathematical theory of epidemics. Proc. Roy. Soc. A {\bf 115} (1927) 700.

\bibitem{Ken57} D. G. Kendall, Discussion of ‘Measles periodicity and community size’ by M. S. Bartlett, J. Roy. Stat. Soc. A {\bf 120} (1957) 
64.
% Kendall provided a spatial generalization of the Kermack and  McKendrick model in a closed population, i.e. neglecting the effects of spatial migration. 

\bibitem{Bar56}  M. S. Bartlett, Deterministic and Stochastic Models for Recurrent Epidemics, 
Berkeley Symp. on Math. Statist. and Prob., Proc. Third Berkeley Symp. on Math. Statist. and Prob., Vol. 4, 81-109 (Univ. of Calif. Press, 1956).

\bibitem{Bar57} M. S. Bartlett, Measles Periodicity and Community Size, 
J. Royal Stat. Soc.  A {\bf 120}, No. 1,  (1957) 48. 
% The phenomenon of recurrent measles epidemics gives no clear evidence of any damping. 
% To investigate the relation time between epidemics and its connection with community size, and the critical size for which this relation applies. 

\bibitem{Fla95} W. D. Flanders and D. G. Kleinbaum, Basic Models for
  Disease Occurrence in Epidemiology, Int. J. Epidemiology {\bf 24} (1) (1995)  1.


\bibitem{Wei13} H. Weiss, The {\small SIR} model and the Foundations of Public Health, MATerials MATematics no. 3 (2013).

 
\bibitem{Cha14} S. Chauhan1, O. P. Misra and J. Dhar. 
Stability analysis of {\small SIR} model with vaccination. 
J.  Comp. and Applied Math. {\bf 4} (1) (2014) 17.

\bibitem{Cha16} D. L. Chao and D. T. Dimitrov, Seasonality and the
  effectiveness of mass vaccination, Math. Biosci. Eng.  {\bf 13(2)} (2016)  249.
  
\bibitem{Rod16} H. S. Rodrigues, Application of {\small SIR} epidemiological
  model: new trends, arXiv:1611.02565 (2016)


\bibitem{Ama20} J. E. Amaro, The D model for deaths by {\small COVID}-19,
  arXiv:2003.13747v1 (2020).

\bibitem{Ama21a} J. E. Amaro, J. Dudouet, and J. N. Orce, 
Global analysis   of the {\small COVID}-19 pandemic using simple epidemiological models, 
Applied Mathematical Modelling {\bf 90} (2021) 995.

  

\bibitem{Coo20} I. Cooper, A. Mondal, and C. G. Antonopoulos,
A {\small SIR} model assumption for the spread of {\small COVID}-19 in different communities
Chaos, Solitons \& Fractals{\bf 139} (2020) 110057.


\bibitem{Kud21} N. A. Kudryashov, M. A. Chmykhov, and M. Vigdorowitsch,
Analytical features of the {\small SIR} model and their applications to {\small COVID}-19,
Applied Mathematical Modelling {\bf 90} (2021) 466. 

\bibitem{Pos20} E. B. Postnikov,
Estimation of {\small COVID}-19 dynamics “on a back-of-envelope”: Does the simplest {\small SIR} model provide quantitative parameters and predictions?
Chaos, Solitons \& Fractals {\bf 135} (2020) 109841.

\bibitem{Fan20} D. Fanelli and F. Piazza, 
Analysis and forecast of {\small COVID}-19 spreading in China, Italy and France, 
Chaos, Solitons \& Fractals {\bf 134} (2020) 109761.

\bibitem{Rad20} A. Radulescu, C. Williams, and K. Cavanagh,
Management strategies in a SEIR-type model of {\small COVID}-19 community spread
Scientific Reports {\bf 10} (2020) 21256. 

\bibitem{Xia21} Xiaowei Chen, Jing Li, Chen Xiao and Peilin Yang,
  Numerical solution and parameter estimation for uncertain {\small SIR} model
  with application to {\small COVID}-19, Fuzzy Optimization and Decision Making
  {\bf 20} (2021) 189. 



\bibitem{Ama21b} J. E. Amaro and J. N. Orce, Monte Carlo simulation of
  {\small COVID}-19 pandemic using statistical physics-inspired probabilities
  (2021) arXiv preprint arXiv:2110.03862


\bibitem{nature1} Gang Xie, 
A novel Monte Carlo simulation procedure for modelling {\small COVID}-19 spread over time, 
Scientific Reports {\bf 10} (2020) 13120. 
    
    
\bibitem{linda}  J. L. S. Allen, 
An introduction to Stochastic Epidemic Models. Pages 81-128 in Fred Brauer, 
Pauline van den Driessche \& Jianhong Wu (Eds.) Mathematical epidemiology, Springer (2008) pp 81.

\bibitem{hakan} H.  Andersson and T.  Britton, 
Stochastic epidemic models and their statistical analysis,  Springer (2000). 



\bibitem{Rod20} W. C. Roda, M. B. Varughese, D. Han, and M. Y. Li, 
 Why is it difficult to accurately predict the {\small COVID}-19 epidemic?
 Infect. Dis. Model. {\bf 5} (2020) 271.

\bibitem{worldometer} https://www.worldometers.info/coronavirus/


\bibitem{Fur20} H.  Furutani, T. Hiroyasu, and Y.  Okuhara,
  Simple method for estimating daily and total {\small COVID}-19 deaths using a
  Gumbel model. Researchsquare DOI: 10.21203/rs.3.rs-120984/v1

\bibitem{Her22} J. Herby, L. Jonung, and S. H. Hanke
A literature review and meta-analysis of the effects of lockdowns on {\small COVID}-19 mortality,
Studies in Applied Economics {\bf 200} (2022) 1.


  
  
\bibitem{Har14} T. Harko, F. S. N. Lobo, and M. K. Mak, 
Exact analytical solutions of the
  Susceptible-Infected-Recovered ({\small SIR}) epidemic model and of the {\small SIR}
  model with equal death and birth rates. 
  Applied Mathematics and Computation {\bf 236}  (2014) 184. 
%   194. Bibcode:2014arXiv1403.2160H. arXiv:1403.2160. doi:10.1016/j.amc.2014.03.030.

\bibitem{Mil12}  J. C. Miller, 
A note on the derivation of   epidemic final sizes. 
Bull. Math. Bio. {\bf 74} (9)  (2012) 2125. 


\bibitem{Die90} O. Diekmann, J. A. P. Heesterbeek, and J. A. J. Metz, 
On the definition and the computation of the basic
  reproduction ratio $R_0$ in models for infectious diseases in
  heterogeneous populations. 
  J. Math. Biol. {\bf 28} (1990) 365.

\bibitem{Gum35} E. J. Gumbel,  "Les valeurs extrêmes des distributions statistiques", Annales de l'Institut Henri Poincaré {\bf 5} (2) (1935) 115; 
J. E. Gumbel, "The return period of flood flows". The Annals of Mathematical Statistics {\bf 12} (1941) 163.

  
\bibitem{Gum54} E. J. Gumbel,  
Statistical theory of extreme values and some practical applications. 
 U.S. Department of Commerce, National Bureau  of Standards. 
 Applied Mathematics  Series. 33 (1st ed.) (1954). 
  
\bibitem{world} https://ourworldindata.org/coronavirus

  
\end{thebibliography}
\end{document}